\documentclass[12pt]{iopart}
\usepackage{graphicx}
\usepackage{iopams}
\usepackage{amssymb}
\usepackage{amscd}
\usepackage{bbm}
\usepackage{epstopdf}
\usepackage{color}
\newcommand{\mycomment}[1]{}

\begin{document}
 
\title{Quadratic corrections to the metaplectic formulation of resonant mode conversion}
\author{A S Richardson$^1$ and E R Tracy$^2$}
\address{$^1$ Plasma Science and Fusion Center, Massachusetts Institute of Technology, Cambridge, Massachusetts, 02139}
\address{$^2$ Department of Physics, College of William and Mary, Williamsburg, Virginia, 23187}

\ead{richardson@psfc.mit.edu}

\date{\today}

\begin{abstract}
The effects of quadratic order terms in the dispersion matrix near a mode conversion are considered.  It is shown that including the corrections due to these quadratic terms gives a better matching between the local solution in the mode conversion region, and the far-field WKB solutions for the incoming and outgoing waves.  This matching is demonstrated by comparison of the asymptotic solution with a numerical solution for a simple one-dimensional conversion.  This procedure for obtaining the corrections due to quadratic order terms can be extended to arbitrary order and, in principle, an outline for performing such an extension is given.
\end{abstract}


\submitto{\JPA}



\section{Introduction and Motivation}

The resonant interaction of linear waves in a nonuniform, or time dependent, medium is a 
phenomenon of great interest in a wide variety of fields.  There is a large
literature going back many years which we will not attempt to survey here
(see, for example, the literature cited in~\cite{PoP}).  In this paper, because the resonance is {\em local},
either in time (for time-dependent backgrounds) or in space (for nonuniform backgrounds), 
we assume that away from the resonance the evolution is well described using
adiabatic methods (e.g. WKB methods).  Hence, the problem reduces to ensuring a good matching between
incoming and outgoing WKB solutions. The resonance is dealt with using an appropriate local
evolution equation that is simpler to solve than the full equation.  Often this is done by simply linearizing
the background dependence of the coupled evolution equations in the immediate vicinity of the
resonance, leading to a Landau-Zener-type model. The local solutions can be expressed in terms
of parabolic cylinder functions which are then matched to the incoming and outgoing WKB solutions. 
This level of approximation to the local dynamics has the defect that the parabolic cylinder 
functions, while capturing the local jump in wave amplitude due to the resonant interaction, fail to capture the slower amplitude variation away from the resonance.  This amplitude variation, which is due to action conservation in a nonuniform or nonstationary background medium, is well modeled by the WKB solutions.  Hence, the asymptotic matching region, where the true solution is well modeled by {\em both} WKB and parabolic cylinder solutions, can be quite small, leading to problems for automated matching procedures.  Here we consider 
how to systematically account for higher order effects in order to improve the robustness of the matching procedure.  Including the quadratic order corrections, for example, dramatically enlarges the matching region.
Quadratic order effects have already been considered by many authors.  We mention, for example, 
Delos and Thorston~\cite{DelosPRL}, Swanson~\cite{Swanson,Swanson2}, and Friedland, {\em et al.}~\cite{FriedlandKaufmanPRL}.
The new contribution of the present paper is that we consider higher order corrections from a 
phase space perspective.  This has the advantage of casting the problem into the simplest possible
form, which is universal in character, while preserving all of the essential features of the particular
problem.  By analogy with problems in particle mechanics, the phase space approach allows a
wide variety of changes of representation for simplifying the problem as much as possible and
cleanly separates the inherent complexity of the problem from complexity which is simply due
to a choice of representation.  While this is worked out in detail only for the case of quadratic order
corrections, in the Appendices we also sketch an argument describing how the results might be
formally extended to arbitrary order.  Another great advantage of phase space methods is that 
they are applicable to multidimensional systems, which is an area of ongoing research.

Of central importance in the phase space theory of mode conversion is the concept of 
a {\em normal form} for the local equation.  This is, in 
a very well-defined sense, the simplest representation of the local dynamics.  The natural physical
interpretation is that it is the representation that puts the `uncoupled' wave operators on the diagonal,
with the off-diagonals giving the coupling.  This change of representation to normal form should be
carried out by an adiabatic transformation, meaning that the polarization basis used to construct
the local representations should be well behaved everywhere and not just
away from the resonance.  Normal forms for one-dimensional conversion, like we
consider here, were studied in great detail by Littlejohn and Flynn \cite{robertandgregannals}. In
dimensions higher than one we mention the work of Littlejohn and Flynn \cite{robertchaos}, 
Braam and Duistermaat 
\cite{braam1,braam2}, Colin de~Verdi\`ere \cite{colindeverdiere1,colindeverdiere2}, 
Kammerer \cite{kammerer} and Tracy and Kaufman \cite{PRL}.

Because the most natural
arena to view WKB theory is phase space, we wish to view conversion as a phase space
phenomenon and develop methods based upon the geometric ideas of Maslov theory.
In previous work 
\cite{PoP,robertandgregannals,Tracy03, Jaunalgorithm}
it was shown that phase space techniques can be used to solve wave problems exhibiting mode conversion between modes of two different polarizations.  
Such multicomponent problems can be written initially in $N\times N$ matrix operator form.  Using the congruent reduction procedure developed by Friedland and Kaufman in 
\cite{allanlazar}, the system can be reduced to a scalar wave equation away from
resonances, leading to traditional WKB methods.  In the vicinity of mode conversion, the 
congruent reduction procedure can be used to reduce the problem to a $2\times 2$ matrix form governing the two interacting modes.  The advantage of the congruent reduction procedure, which 
uses general congruence transformations rather than unitary transformations, is that the change 
of polarization basis used to carry it out can be smooth everywhere.  Diagonalization using unitary operators does not have this property due to the near-denegeracy of eigenvalues, which is the essential characteristic
of mode conversion.  

As an example of a physical problem which exhibits mode conversion, consider the equations for the propagation of an electromagnetic wave in a cold plasma.  From Maxwell's equations we have
\begin{equation}\label{eq:maxwell}
\nabla \times \nabla \times \bi{E} - \frac{\omega^2}{c^2} \mathbf{K}\cdot \bi{E} = 0.
\end{equation}
The dielectric tensor $\mathbf{K}$ for a cold plasma can be shown to be \cite{stix:waves}
\begin{equation}
\mathbf{K} = 
\left(\begin{array}{ccc}S & -iD & 0 \\iD & S & 0 \\0 & 0 & P\end{array}\right),
\end{equation}
where the functions $S$, $D$, and $P$ depend on the density of the plasma, and the background magnetic field.  Spatial variations in these plasma parameters lead to spatial variations in the nature of propagating solutions, and in certain conditions \cite{Jaunalgorithm}, this problem will exhibit mode conversion. 

Returning to a generic mode conversion problem, we consider the problem at a given frequency, and write the two equations for the coupled wave modes together,
\begin{equation}\label{eq:wave}
\mathbf{\hat D}(x,-i\partial_x;\omega) \cdot \Psi(x) = 0 .
\end{equation}
While this equation works for waves in multiple spatial dimensions, in this paper we limit our analysis to the case where $x$ is one-dimensional.  Additionally, we will suppress the frequency dependence for brevity of notation.  Using the symbol calculus \cite{Tracy02,Littlejohn}, we can define the symbol of the wave operator as a matrix-valued function on wave phase space, $\mathbf{D}(x,k)$.  Here, the variable $k$ corresponds to the operator $-i\partial_x$, and products of conjugate variables correspond to symmetrized operators (e.g., the symbol $xk$ maps to the operator $-i(x \partial_x+\partial_x x)/2$).  

\begin{figure}
\begin{center}
\includegraphics[scale=1]{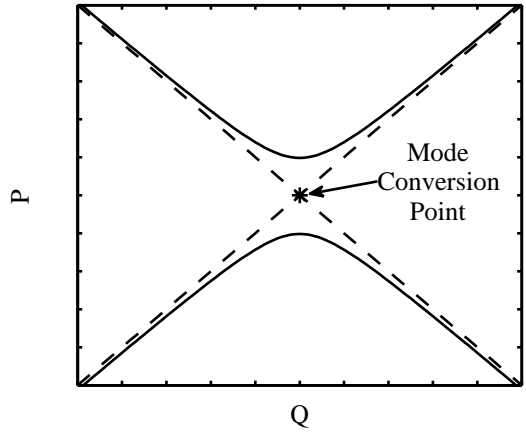}
\end{center}
\caption{\label{fig:crossing}
The phase space structure of a typical ``avoided crossing'' mode conversion.  The hyperbolic dispersion curves are the solid lines, and the dashed lines are the dispersion curves for the uncoupled modes.
}
\end{figure}

In the vicinity of a mode conversion, there are two roots of the dispersion relation $\det(\mathbf{D}(x,k)) = 0$.  These two curves in phase space locally have a hyperbolic structure (an ``avoided crossing'', see Figure (\ref{fig:crossing})).  Linearizing the $x$ and $k$ dependence of the dispersion matrix about the center of the hyperbola, and then converting this linearized symbol back to an operator, gives a set of coupled equations which can be solved for the local wave fields.  Matching these local solutions onto uncoupled WKB solutions (which are a good approximation to the solutions far from the mode conversion region) gives transmission and conversion coefficients for the incoming and outgoing waves.  These coefficients can be used to treat the mode conversion as a ray-splitting process, where the incoming ray is split into two outgoing rays, one for each mode.  

\begin{figure}
\begin{center}
\includegraphics[scale=0.5]{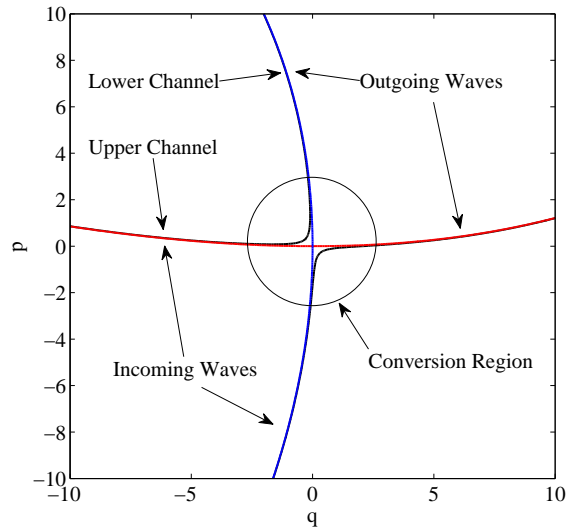}
\end{center}
\caption{\label{fig:uncoupled_modes}
Dispersion surfaces for the uncoupled WKB modes, defined by $D_{11}(q,p)=0$ and $D_{22}(q,p)=0$, cross at the mode conversion point.  The coupled dispersion surface, defined by solving $\det (\mathbf{D}(q,p))=0$, has a hyperbolic structure in the mode conversion region.  Its branches asymptote to the uncoupled dispersion surfaces.  Quadratic terms in the expansion of $D_{11}(q,p)$ and $D_{22}(q,p)$ about the mode conversion point cause the dispersion surfaces to curve.  The coordinates $(q,p)$ are described in Section \ref{sec:coords}. 
}
\end{figure}

This ray-splitting approach captures the jump in amplitude caused by the coupling between the two modes at linear order in phase space variables.  However, higher order terms in the wave equation can lead to additional effects.  For example, the amplitude variation familiar from WKB theory is not captured by the linear order solution.  This could cause difficulties when attempting to match the local wave fields onto the incoming and outgoing WKB solutions.  Figure (\ref{fig:compare_linear}) illustrates this effect by comparing numerical solutions to the linear order solutions.  In this example, the local solution captures the jump in amplitude at the mode conversion, but misses the slow WKB amplitude variation.  This limits the matching region to a small range right near the mode conversion, which could make numerical ray-tracing algorithms somewhat unstable.  However, this example also suggests that it may be possible to calculate the local solution and include the effects of the higher order terms.  As will be shown, one of these effects is an amplitude variation which matches the slow WKB amplitude variation.

In this paper we will first briefly review the linearization procedure which leads to the parabolic cylinder functions as local solutions in the mode conversion region.  We will then consider the effect of adding generic quadratic terms to the wave equation.  These terms will modify the uncoupled dispersion relations (which are given by the diagonal elements of the dispersion matrix, after it has been transformed into a ``normal form''), which changes the far-field WKB solutions for the incoming and outgoing waves.  When the new quadratic terms are added to the coupled equations, they will give us a new local solution for the wave fields.  This local solution will contain both non-propagating ``near-field'' contributions and propagating contributions to the original, linear order, local solution.  The near-field terms do not propagate, and therefore do not affect the matching.  The other modifications, however, are phase and amplitude corrections which make the local solution better match the far-field WKB solutions.  We find that the quadratic order terms do not modify the $S$-matrix (WKB connection) coefficients.  
Last, we give a comparison with numerical solutions for a simple example.

There are three appendices to this paper.  In \ref{sec:nf} we show how to transform the dispersion matrix for the one-dimensional problem into normal form, through second order in phase space variables. In  \ref{sec:extension}, we argue that the calculations of this paper can be extended to higher order in the phase space variables, since it is possible to transform the dispersion matrix into normal form at any order. Finally, in \ref{sec:moyal}, we calculate corrections to the normal form transformation which are due to the Moyal star product.  We show that, while these effects can be calculated, they introduce a higher order correction, and therefore can be neglected at the quadratic order we consider in this paper.

\section{Qualitative Discussion of Results}

There are three primary results presented in this paper.  These are
\begin{enumerate}
\item New local solutions are found, which include the effects of quadratic order terms in the dispersion matrix
\item The matching region is expanded,
\item The transmission the conversion coefficients are unchanged through $\mathcal{O}(\epsilon)$
where $\epsilon$ is a formal small parameter associated with the corrections.
\end{enumerate}

For the first result, we Taylor expand the dispersion matrix out to second order in phase space variables, and truncate.  We then choose a particular representation, and use it to convert the truncated dispersion matrix into a pair of coupled equations for the fields.  This approximate wave equation is valid near the mode conversion, and solutions to this equation are new local solutions.  The new local solutions have the form of the local solution obtained from truncation of the dispersion matrix at linear order, but with corrections due to the quadratic order terms.

\begin{figure}
\begin{center}
\includegraphics[scale=0.9]{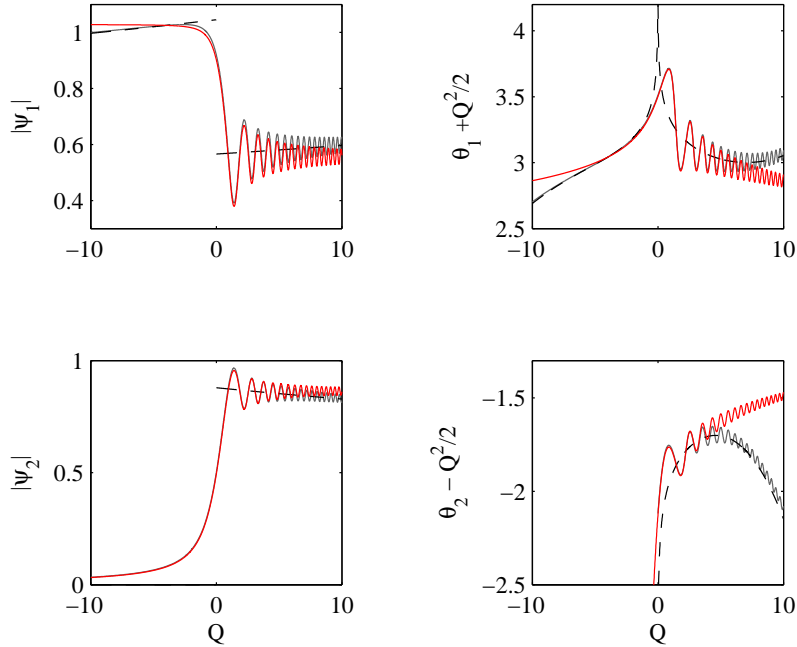}
\end{center}
\caption{\label{fig:compare_linear}
Comparison of WKB (black dashed), numerical (gray), and linear order local (red) solutions.
}
\end{figure}

\begin{figure}
\begin{center}
\includegraphics[scale=0.9]{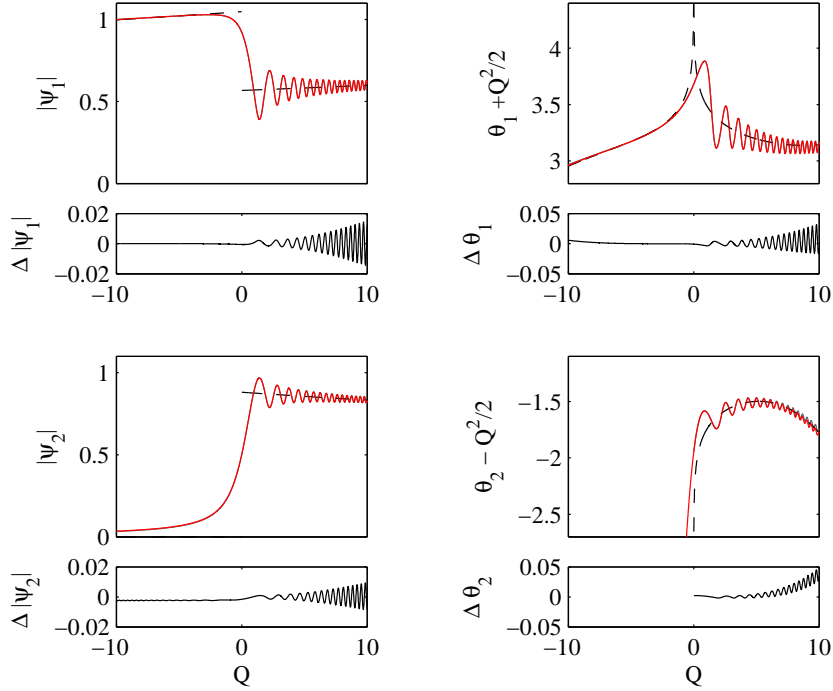}
\end{center}
\caption{\label{fig:compare_quad}
Comparison of WKB (black dashed), numerical (gray), and quadratic order local (red) solutions.  Since the numerical solutions are almost entirely covered by the local solutions, the subfigures show the difference between the numerical and the quadratic order local solutions.  The parameters for this simulation are $\epsilon = 1$, $a_1=3\times 10^{-3}$, $b_1 = 7.44\times 10^{-3}$, $c_1 = 1\times 10^{-4}$, $a_2=3\times 10^{-3}$, $b_2 = 7.44\times 10^{-3}$, and $c_2 = -5\times 10^{-4}$.
}
\end{figure}

The expansion of the matching region can be seen most clearly in the comparison of the new local fields to the WKB approximations which are valid far from the mode conversion.  In Figure (\ref{fig:compare_linear}) the linear order local solutions are plotted along with the WKB approximations and ``exact'' numerical solutions.  The local solutions can be used for matching, but higher order terms in the equations result in amplitude and phase variations in the WKB solutions that are not in the local solutions.  This means the matching must be done close the the mode conversion point in order to obtain reasonable results.  In Figure (\ref{fig:compare_quad}), the corrected local solutions are plotted instead of the linear order solutions.  The quadratic order corrections clearly make the new local solutions match the numerical and WKB solutions over a much larger range.  This means that the region in which the matching can be accurately performed has now been greatly expanded.

That the transmission and conversion coefficients are unchanged by the second order terms will be seen in Section \ref{sec:2nd_matching}, where the functional forms of the new local solutions and the WKB solutions are matched asymptotically.  The implication of this result is that the previously obtained transmission and conversion coefficients can still be used in analysis or numerical routines, without modifications due to quadratic order terms.

A further result which will be discussed in \ref{sec:extension} is that the ideas and analysis used in this paper can be used to obtain further corrections to the local fields, order by order in the phase space variables.  The Taylor's series for the dispersion matrix can be can be truncated at some order, and then thrown into normal form.  This new approximation to the dispersion matrix can then be used to write equations for the local fields.  Solutions to these equations will give new expressions for the local fields, which will contain corrections out to the order where the series was truncated.

\section{Linearization of the coupled system\label{sec:lin}}

This review of the solution of the mode conversion problem follows the solution described in Reference \cite{PoP}.  We give this review of the solution in some detail because the solution to the quadratic order problem proceeds along the same lines.  The first step of the solution is to linearize the symbol of the wave operator about the mode conversion point.  Then, transform the linearized symbol via a change of polarization basis and a linear canonical transformation of phase space, in order to simplify the symbol as much as possible.  Convert this approximate symbol back into an operator, which gives a new ``local'' wave equation which can be solved analytically.  The solutions can then be analyzed in various representations, which correspond to different choices of coordinates in phase space.  The transmission and reflection coefficients are then derived by matching the local solution onto the far-field WKB solutions.

\subsection{Coordinates Used in this Calculation\label{sec:coords}}

\begin{figure}
\begin{center}
\begin{tabular}{ccc}
\includegraphics[scale=0.8]{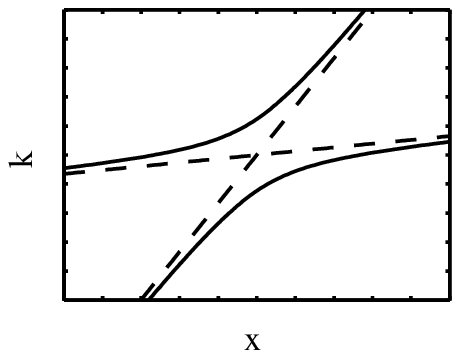} &
\includegraphics[scale=0.8]{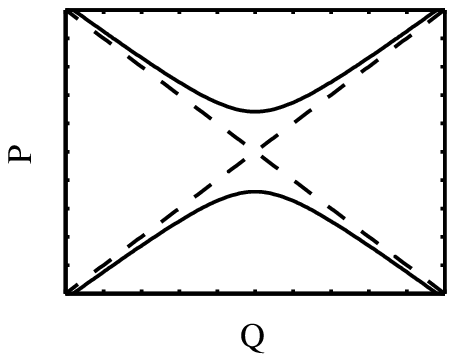} &
\includegraphics[scale=0.8]{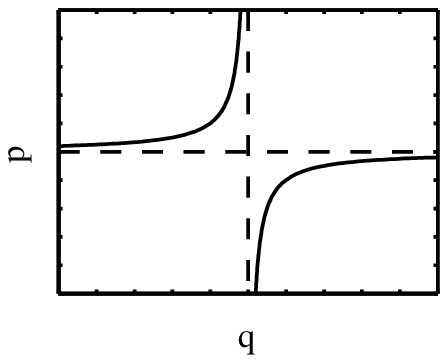} \\
~~(a) & ~~~(b) & ~~~(c)
\end{tabular}
\end{center}
\caption{\label{fig:coords_physical}
The dispersion curves in three different coordinate systems: (a) the physical coordinates $(x,k)$, (b) the symmetrized coordinates $(Q,P)$ of Flynn and Littlejohn \cite{robertandgregannals}, and (c) the coordinates $(q,p)$ of Tracy and Kaufman \cite{Tracy02}.
}
\end{figure}

By performing this calculation in phase space, the underlying symmetry between configuration coordinates and wave numbers becomes evident, and the techniques of linear canonical transformations can be used to simplify the problem.  This can lead to notational difficulties, however, since any particular choice of linear canonical transformation of the phase space variables can lead to a new notation for the variable's names.  In this paper, there will be primarily three coordinate systems used to label the phase space variables, as illustrated in Figure (\ref{fig:coords_physical}).  The first set of coordinates, $(x,k)$, is used for the physical coordinates in which the problem is naturally written.  In these coordinates, the hyperbolic structure of the dispersion surfaces near a mode conversion will be oriented at some arbitrary angle in phase space [see Figure (\ref{fig:coords_physical}a)].  The second set of coordinates, denoted $(Q,P)$, is the pair of symmetrized coordinates used by Flynn and Littlejohn in \cite{robertandgregannals}.  In these coordinates the asymptotes of the hyperbolic avoided crossing are oriented on the diagonals [see Figure (\ref{fig:coords_physical}b)].  Also, these variables have been normalized so that the diagonal elements of the linearized dispersion matrix form a canonical pair.  The final set of coordinates is that used by Tracy and Kaufman in \cite{Tracy02}.  In these coordinates, denoted $(q,p)$, the axes are the asymptotes of the avoided crossing [see Figure (\ref{fig:coords_physical}c)].  The diagonal elements of the linearized dispersion matrix are again normalized to form a canonical pair.  An advantage of this choice of coordinates is that it puts the linearized dispersion matrix into a kind of normal form, where the two diagonal elements take the simple form $D_{11}(q,p) = -p$ and $D_{22}(q,p) = q$.  As we will see, the equations for the local fields are particularly simple to solve in these coordinates.

Associated with each of these three different choices of coordinates is a different representation of the fields.  The linear canonical transformation from one pair of coordinates to another induces a metaplectic transformation of the fields (which can be thought of as a generalization of the Fourier transform).  For example, the linear canonical transformation from the variables $(q,p)$ to $(Q,P)$ is 
\begin{equation} \label{eq:Q_rep_transformation}
\left(
\begin{array}{c}
Q \\
P
\end{array}
\right) = \frac{1}{\sqrt 2}
\left(
\begin{array}{cc}
 1 & 1 \\
 -1 & 1
\end{array}
\right)
\left(
\begin{array}{c}
 q \\
 p
\end{array}
\right).
\end{equation}
The associated metaplectic transformation is
\begin{equation}\label{eq:Q_rep_meta}
\Psi(Q) = \int e^{iF_1(Q,q)} \Psi(q) \, dq ,
\end{equation}
where $F_1(Q,q)$ is the mixed-variable generating function for the canonical transformation:
\begin{equation}
F_1(Q,q) = \frac{1}{2} (Q^2 - 2\sqrt{2} Q q + q^2) .
\end{equation}
These equations illustrate the relationship of the metaplectic transformation to the Fourier transform.  If the transformation matrix in Equation (\ref{eq:Q_rep_transformation}) is squared, it would give a rotation by $\pi/2$ in phase space, and the metaplectic transformation in Equation (\ref{eq:Q_rep_meta}) 
(iterated twice ) would become a Fourier transformation.  So, Equation (\ref{eq:Q_rep_meta}) can be 
thought of as the ``square root'' of a Fourier transformation.

\subsection{Linearization and Solution}

The Taylor's series expansion of the symbol of $\mathbf{ \hat D}$ about the mode conversion point $(x_*,k_*)$ is given by

\begin{eqnarray}\label{eq:taylor_matrix}
\mathbf{D}(x,k) &= \mathbf{D}(x_*,k_*) 
+ (x-x_*)\frac{\partial \mathbf D}{\partial x}  
+ (k-k_*)\frac{\partial \mathbf D}{\partial k}  + 
 \frac{(k-k_*)^2}{2}\frac{\partial ^2\mathbf D}{\partial k^2} \ldots
\end{eqnarray}

where all derivatives are evaluated at the conversion point $(x_*,k_*)$.
First shift the origin in phase space to $(x_*,k_*)$.  The shift of origin in $x$ is performed by the change of variables $x \rightarrow x-x_*$;
\begin{eqnarray}
\psi'(x) = \psi(x-x_*).
\end{eqnarray}
The shift in $k$ is accomplished by multiplication by a phase:
\begin{eqnarray}
\psi''(x) = e^{ik_* x}\psi(x).
\end{eqnarray}
That this is a shift in $k$ can be seen by computing the Fourier transform of $\psi''(x)$:
\begin{eqnarray}
[\mathcal{F}\psi''](k) = \int e^{-ikx} \psi''(x) \, dx = \int e^{-ikx + ik_* x} \psi(x) \, dx = [\mathcal{F}\psi](k-k_*).
\end{eqnarray}

Now truncate the series in Equation (\ref{eq:taylor_matrix}) at linear order in phase space variables.  As shown in Reference \cite{robertandgregannals}, the resulting matrix can be simplified by choosing the polarization basis which makes its off-diagonal elements constant:
\begin{eqnarray}
\mathbf{\widetilde D}(x,k) = 
\left(
\begin{array}{cc}
D_{11}(x,k) & \eta \\
\eta^* & D_{22}(x,k)
\end{array}
\right) .
\end{eqnarray}
We now have the situation shown in Figure (\ref{fig:coords_physical}a).  The dispersion surface
$det(\tilde{\bf D})(x,k)=0$ has a locally hyperbolic structure, and the center of the hyperbola is at the 
origin in phase space.  The asymptotes of the hyperbola are given by the zeros of the diagonal elements of $\mathbf{\widetilde D}(x,k)$.  These diagonal elements can be simplified by performing a linear canonical transformation of the phase space coordinates $(x,k)$ to the coordinates $(q,p)$ of Figure (\ref{fig:coords_physical}c).  This puts the asymptotes of the hyperbola on the $q$ and $p$ axes.  With this choice of linear canonical transformation, the linearized dispersion matrix becomes
\begin{equation}\label{eq:linearized}
\mathbf{\widetilde D}(q,p) = 
\left(
\begin{array}{cc}
-p & \tilde\eta \\
\tilde\eta^* & q
\end{array}
\right) ,
\end{equation}
where the coupling $\tilde\eta$ includes the Poisson bracket of the diagonals as a normalization;
\begin{eqnarray}
\tilde \eta = \frac{\eta}{ \{ D_{11},D_{22}\}\vert_{x=0,k=0}}.
\end{eqnarray}
The complex constant $\tilde\eta$ is the coupling between the two modes.  In the limit $\tilde\eta \to 0$, the two modes would become uncoupled.  The dispersion relations for the two coupled modes are given by setting the eigenvalues of the dispersion matrix equal to zero.  For $\tilde\eta \to 0$, the polarization basis chosen in Equation (\ref{eq:linearized}) diagonalizes the dispersion matrix, and therefore the diagonal elements are the dispersion functions for the uncoupled modes.  For nonzero $\tilde\eta$ the diagonals are no longer equal to the eigenvalues.  However, far from the mode conversion region the diagonals are approximately equal to the eigenvalues, and so provide a good approximation to the dispersion functions for the two modes.  This means that the WKB method can be used to construct approximate solutions, using either the diagonal elements of the dispersion matrix (in the ``normal form'' given by Equation (\ref{eq:linearized})) or the eigenvalues of the dispersion matrix.  We refer to these two types of approximate solutions as the uncoupled and coupled WKB solutions, respectively.  

The approximate dispersion matrix in Equation (\ref{eq:linearized}) can now be converted back from a symbol into an operator, and we have an equation for the local wave field in the $q$ representation,
\begin{equation}\label{eq:local_lin_equation}
\left(
\begin{array}{cc}
i\partial_{q} & \tilde\eta \\
\tilde\eta^* & q
\end{array}
\right)
\left(
\begin{array}{c}
 \psi_1(q) \\
\psi_2(q)
\end{array}
\right) = 0 .
\end{equation}
Notice that the second row of this matrix equation is now an algebraic equation instead of a differential equation.  Solving this algebraic equation for $\psi_2(q)$ in terms of $\psi_1(q)$ and inserting the resulting expression into the first row gives a first-order differential equation for $\psi_1(q)$.  Integrating this equation, we obtain the following solution:
\begin{equation}
\Psi(q)=\left(
\begin{array}{c}
A e^{-i\vert\tilde\eta\vert^2 \ln q} \\
-\frac{A \tilde\eta^*}{q} e^{-i\vert\tilde\eta\vert^2 \ln q}
\end{array}
\right)=\left(
\begin{array}{c}
A q^{-i\vert\tilde\eta\vert^2} \\
-A \tilde\eta^* q^{-i\vert\tilde\eta\vert^2 -1}
\end{array}
\right) ,
\end{equation}
where $A$ is a constant of integration.  This expression must be evaluated with care for negative values of $q$ because of the branch in the complex logarithm.  If we take the branch to be just below the negative axis, then $\ln(q) = \ln(|q|) +i\pi$ and the solution can be written in terms of the absolute value of $q$:
\begin{equation}\label{eq:local_linear_solution}
\Psi(q)=\left(
\begin{array}{c}
 \psi_1^{(0)}(q) \\
\psi_2^{(0)}(q)
\end{array}
\right) = 
\left(
\begin{array}{c}
A \vert q \vert^{-i\vert\tilde\eta\vert^2 } f(q) \\
-\frac{\tilde\eta^* A}{q} \vert q \vert^{-i\vert\tilde\eta\vert^2} f(q)
\end{array}
\right) ,
\end{equation}
where the function $f(q)$ is defined for real $q$ as
\begin{equation}\label{eq:local_linear_solution_factor}
f(q)=
\cases{1/\tau &for $q<0$ \\
1 &for $q>0$}
\end{equation}
in order to deal with the singularity at $q=0$.  The variable $\tau=e^{-\pi \vert\tilde\eta\vert^2}$ is the transmission coefficient.  If we absorb the factor of $1/\tau$ into the definition of $A$, then we can see that the absolute value of $\psi^{(0)}_1$ decreases by a factor of $\tau$ as we go from negative $q$ to positive $q$.  For this solution, the energy lost from the upper channel is converted into energy outgoing in the lower channel.

The solution given in Equations (\ref{eq:local_linear_solution}) and (\ref{eq:local_linear_solution_factor}) is only one of two possible types of solutions for the local fields.  The other solution can be found by placing the branch cut just above the negative $q$ axis, instead of below it.  In this case, $\ln(q) = \ln(|q|) -i\pi$ for $q < 0$.  This means that now the amplitude of $\psi^{(0)}_1$ increases by a factor of $1/\tau$ as we go from negative to positive $q$.  This solution corresponds to the case where energy is coming in on both the upper and lower channel, but the amplitudes and phases are just right so that all of the energy leaves in the upper channel.

\subsection{Alternative Representations\label{sec:alt_rep}} 

The form of the solution given in Equation (\ref{eq:local_linear_solution}) is particularly convenient for solving the linearized system of equations, and calculating the transmission and conversion coefficients.  However, the coordinates $(q,p)$ of Equation (\ref{eq:local_linear_solution}) are related to the physical coordinates $(x,k)$ by some linear canonical transformation.  Therefore, to find the solution in the $x$ representation requires computing a  metaplectic transformation of the solution $\Psi(q)$.  An additional inconvenience of using the $q$ representation is that the solution for the lower channel has a singularity at $q=0$.  While this is not too surprising since the dispersion manifold for this mode is the $p$ axis, it makes analysis of this function tricky.

As an example of how to transform the solution in Equation (\ref{eq:local_linear_solution}) into a different representation, we will transform to the variables $(Q,P)$ described in Section \ref{sec:coords}.  This transformation induces the metaplectic transformation given in Equation (\ref{eq:Q_rep_meta}).  A table of integrals or a computer algebra system like \sf Maple \rm can be used to evaluate this integral, resulting in the $Q$ representation of the upper channel being written in terms of the parabolic cylinder function $U(a,Q)$:
\begin{eqnarray} \label{eq:psi1_0_Q_int}
\psi^{(0)}_1(Q) &= \int e^{iF_1(Q,q)} \psi^{(0)}_1(q) \, dq  \\
&= \tilde A e^{3\pi i/4} \; U\left(i|\tilde\eta|^2 -1/2, -(1+i)Q \right) .
\end{eqnarray}
Here, $\tilde A$ is a complex amplitude, whose value is set when we match this local function to the incoming wave.   It is related to the amplitude $A$ of Equation (\ref{eq:local_linear_solution}) by a constant factor; $\tilde A = -i \sqrt{2 \pi} e^{3\pi |\tilde\eta|^2/4} A$.

Because of the singular nature of $\psi^{(0)}_2(q)$, the metaplectic integrals to convert this to a different representation become difficult to evaluate.  In \cite{PoP}, Tracy et al.\ compute the Fourier transform of $\psi^{(0)}_2(q)$ using contour integrals, which gives the $p$ representation of the lower channel.  An alternative approach is to use the $Q$ representation of the wave equation to write $\psi^{(0)}_2(Q)$ in terms of the parabolic cylinder function and its derivatives.  
\begin{eqnarray}
\psi^{(0)}_2(Q)  &= \int e^{iF_1(Q,q)} \left( \frac{-i \partial_q}{\tilde \eta} \,\psi^{(0)}_1(q) \right)\, dq  \\
&=\frac{1}{\tilde \eta}\int \left(i \partial_q e^{iF_1(Q,q)} \right)   \psi^{(0)}_1(q)\, dq \\
&=\frac{1}{\tilde \eta}\int \left(\frac{1}{\sqrt{2}}\left( Q -i\partial_Q \right) e^{iF_1(Q,q)} \right)   \psi^{(0)}_1(q)\, dq \\
&= \frac{1}{\tilde\eta \sqrt 2} (Q -i\partial_Q) \psi^{(0)}_1(Q) \\
&= - \tilde A \tilde\eta^*   \,U\left(i|\tilde\eta|^2 +1/2, -(1+i)Q \right)  .
\end{eqnarray}
Recurrence relations for the parabolic cylinder function provide the simplification to the last line above.
This representation of the solution can be easily compared to numerical simulations of the original system of equations, since methods for calculating the parabolic cylinder functions are readily available.  See Figure (\ref{fig:compare_linear}).

\subsection{Transmission and Conversion Coefficients\label{sec:coeffs}}

The transmission and conversion coefficients for this problem are derived from asymptotically matching the local solution in Equation (\ref{eq:local_linear_solution}) to the far-field WKB solutions, which are defined using the eigenvalues of the dispersion matrix as the ray hamiltonians.  Because of the rapid variation of the eigenvectors in the mode conversion region, the WKB solution breaks down there.  However, there is a region where both the local and WKB solutions are valid, and have the same functional form.  By matching the solutions in this region, a globally valid approximate solution can be formed.

The eigenvalues of the dispersion matrix in Equation (\ref{eq:linearized}) are 
\begin{eqnarray}
\lambda_\pm=\frac{q-p}{2} \pm \sqrt{\left(\frac{q+p}{2}\right)^2 +|\tilde\eta|^2}\,.
\end{eqnarray}
The choice of which eigenvalue to consider is determined by the sign of $q$.  For $q<0$, there is a solution to $\lambda_+ = 0$, while for $q>0$, a zero of the other eigenvalue $\lambda_-$ exists 
[see Fig.~(\ref{fig:coords_physical}(c))].

Setting $\lambda =0$ and solving for $p(q)$ gives the WKB phase:
\begin{eqnarray}
\Theta(q) = \int^q p(q') \, dq = \int^q \left(\frac{-|\tilde\eta|^2}{q'}\right) \, dq'
=-|\tilde\eta|^2 \ln(q).
\end{eqnarray}
The WKB amplitude is found by taking the $p$ derivative of the eigenvalue:
\begin{eqnarray}
A(q) = \left| \frac{\partial \lambda_\alpha}{\partial p}\right|^{-1/2}_{p=-|\tilde\eta|^2/q}.
\end{eqnarray}
Evaluation of the $p$ derivative, and simplification of the result, gives
\begin{eqnarray}
A(q) &= \left|-1 + \frac{|\tilde\eta|^2}{q^2} + \mathcal{O}\left(\frac{|\tilde\eta|^4}{q^4}\right) \right|^{-1/2} \\
&= \left|1 + \frac{|\tilde\eta|^2}{2 q^2} + \mathcal{O}\left(\frac{|\tilde\eta|^4}{q^4}\right) \right|, \label{eq:linear_wkb_amp}
\end{eqnarray}
where we have used the fact that we are interested in the eigenvalue $\lambda_\alpha = \lambda_{-{\rm sgn}(q)}$.  Together, the WKB amplitude and phase give the WKB solution for the upper channel:
\begin{eqnarray}
\psi_1^{{\rm WKB}}(q) = e^{-i|\tilde\eta|^2 \ln(q)} \left|1 + \frac{|\tilde\eta|^2}{2 q^2} + \mathcal{O}\left(\frac{|\tilde\eta|^4}{q^4}\right) \right|.
\end{eqnarray}
The transmission coefficient is obtained by matching this WKB solution to the local solution.  However, far from the mode conversion, where the WKB approximation is valid and the matching can be performed, we have $q^2 \gg |\tilde\eta|^2$.  This means that the $\mathcal{O}\left(\frac{|\tilde\eta|^2}{q^2}\right)$ term can be dropped from the amplitude.  We are then left with an expression that has the same functional form as the local solution given in Equation (\ref{eq:local_linear_solution}).  This confirms our previous identification of $\tau=e^{-\pi|\tilde\eta|^2}$ as the transmission coefficient.

\mycomment{%
However, the transmission coefficient for the upper channel can be more quickly derived using the uncoupled WKB solution, which is defined using the diagonal element $D_{11} = -p$ as the ray hamiltonian.  In this case, the WKB solutions are constant functions of $q$.  Matching the local solution from Equation (\ref{eq:local_linear_solution}) to constant functions at the points $q = -q_M$ and $q = +q_M$ gives results which depend on the chosen matching points $\pm q_M$;
\begin{eqnarray}
\psi^{{\rm WKB}}_{{\rm left}}(q) &\equiv \psi_1^{(0)}(-q_M) = Ae^{\pi\vert\tilde\eta\vert^2} e^{-i\vert\tilde\eta\vert^2 \ln |q_M|}\\
\psi^{{\rm WKB}}_{{\rm right}}(q) &\equiv \psi_1^{(0)}(-q_M) = A e^{-i\vert\tilde\eta\vert^2 \ln |q_M|}.
\end{eqnarray}
Even though these depend on the matching point, we can derive a transmission coefficient which is independent of the matching point by taking the ratio of these two;
\begin{eqnarray}
\tau \equiv \frac{\psi^{{\rm WKB}}_{{\rm right}}(q)}{\psi^{{\rm WKB}}_{{\rm left}}(q)} = e^{-\pi\vert\tilde\eta\vert^2}.
\end{eqnarray}
This is the same result that would be obtained by using the coupled WKB solutions to do the matching.
}

The derivation of the conversion coefficient is complicated by the need to evaluate the lower channel in the $p$ representation.  The integral in the Fourier transform which changes $\psi^{(0)}_2(q)$ into the $p$ representation must be treated as a contour integral because of the singularity at $q=0$.  The details of this calculation are given in Reference \cite{PoP}, and the result is that $\psi^{(0)}_2(p)$ is zero for negative $p$, while for positive $p$ it is 
\begin{eqnarray}\label{eq:psi_2_0_p_rep}
\psi^{(0)}_2(p) = \frac{A}{\tau} \frac{\sqrt{2\pi \tau i}}{\tilde\eta \Gamma(i|\tilde\eta|^2)}  p^{i|\tilde\eta|^2}.
\end{eqnarray}
Comparing this to $\psi_1^{(0)}(q)$ for negative $q$ we find that the conversion coefficient is
\begin{eqnarray}\label{eq:con_coeff}
	\beta^* \equiv \frac{\sqrt{2\pi \tau i}}{\tilde\eta \Gamma(i|\tilde\eta|^2)}.
\end{eqnarray}
The incoming and outgoing amplitudes are therefore related by a scattering matrix:
\begin{eqnarray}
	\left(\begin{array}{c}
	\overline{\psi_1^{(0)}}(+q) \\
	\overline{\psi_2^{(0)}}(+p)
	\end{array} \right)
	 = \left(\begin{array}{cc}
	\tau & -\beta \\
	\beta^* & \tau 
	\end{array} \right)
	\left(\begin{array}{c}
	\overline{\psi_1^{(0)}}(-q) \\
	\overline{\psi_2^{(0)}}(-p)
	\end{array} \right),
\end{eqnarray}
where the amplitude functions are defined as in Reference \cite{Tracy02}:
\begin{eqnarray}
\overline{\psi_1^{(0)}}(\pm q) &\equiv |q|^{i|\tilde\eta|^2}\psi_1^{(0)}(\pm q)\\
\overline{\psi_2^{(0)}}(\pm p) &\equiv |p|^{-i|\tilde\eta|^2}\psi_2^{(0)}(\pm p).
\end{eqnarray}

\section{Extension to Higher Order}

In general, the Taylor's series expansion of the dispersion matrix in Equation (\ref{eq:taylor_matrix}) will contain terms of higher order than the linear terms kept in the approximation of the previous section.  
For example, second order derivatives, such as the curl of the curl in Equation (\ref{eq:maxwell}), will lead to terms quadratic in $p$ in the dispersion matrix.
Such higher order terms will have an effect on both the far-field WKB solutions, and on the local solution.  If we can calculate these effects, then we can obtain a better match of the local solution to the incoming and outgoing WKB solutions.  This will allow us calculate any corrections that there may be to the scattering coefficients.

As shown in \ref{sec:nf}, the quadratic approximation to the dispersion matrix can be put into the normal form
\begin{equation}\label{eq:disp_matrix}
\mathbf{D}(q,p)=
\left(
\begin{array}{cc}
D_{11}(q,p) &  \tilde\eta  \\
 \tilde\eta^* & D_{22}(q,p)
\end{array}
\right).
\end{equation}
The diagonal elements of this normal form matrix include arbitrary quadratic terms in addition to the linear terms which were described previously, while the off-diagonal terms are constant through quadratic order in the phase space variables.  Introducing the small parameter $\epsilon$ to keep track of the quadratic terms we obtain
\begin{eqnarray}
D_{11}(q,p) = -p + \epsilon (a_1 p^2 + b_1 pq + c_1 q^2) 
\end{eqnarray}
and
\begin{eqnarray}
D_{22}(q,p) = q   + \epsilon  (c_2 p^2 + b_2 pq + a_2 q^2) .
\end{eqnarray}

This second order matrix valued function on phase space can now be converted into a pair of coupled second order differential equations for the two modes, in the $q$ representation:  
\begin{equation}\label{eq:2nd_order_wave_eqn}
\left(
\begin{array}{cc}
\widehat D_{11} &  \tilde\eta  \\
 \tilde\eta^* & \widehat D_{22}
\end{array}
\right)
\left(
\begin{array}{c}
\psi_1(q) \\
\psi_2(q)
\end{array}
\right) = 0,
\end{equation}
where the operators on the diagonals are 
\begin{eqnarray}
\widehat D_{11} = i\partial_q + \epsilon (a_1 (-i\partial_q)^2 -\frac{i b_1}{2} (\partial_q q + q\partial_q) + c_1 q^2),
\end{eqnarray}
and
\begin{eqnarray}
\widehat D_{22} = q   + \epsilon  (c_2 (-i\partial_q)^2 -\frac{i b_2}{2} (\partial_q q + q\partial_q)  + a_2 q^2).
\end{eqnarray}
As seen in the previous section, the lowest order solution for the second channel has a singularity at the origin when it is written in the $q$-representation, and therefore was Fourier transformed to obtain the $p$-representation given in Equation (\ref{eq:psi_2_0_p_rep}).  Anticipating that this form of solution will persist even in the presence of the new $\mathcal{O}(\epsilon)$ terms, we will analyze the lower channel in Fourier space.  The Fourier transform of these equations gives the $p$-representation of the equations:
\begin{equation}\label{eq:2nd_order_wave_eqn_p_rep}
\left(
\begin{array}{cc}
\widehat D'_{11} &  \tilde\eta  \\
 \tilde\eta^* & \widehat D'_{22} 
\end{array}
\right)
\left(
\begin{array}{c}
\psi_1(p) \\
\psi_2(p)
\end{array}
\right) = 0,
\end{equation}
which has diagonal elements
\begin{eqnarray}
\widehat D'_{11} = -p + \epsilon (a_1 p^2 +\frac{i b_1}{2} (p\partial_p + \partial_p p) + c_1 (i\partial_p)^2),
\end{eqnarray}
and
\begin{eqnarray}
\widehat D'_{22} =  i\partial_p + \epsilon(a_2(i\partial_p)^2 +\frac{i b_2}{2}(p\partial_p + \partial_p p) +c_2 p^2 ).
\end{eqnarray}

We will proceed by first finding the new far-field WKB solutions, and then finding the coupled local solution.  Then we will examine the local solution and match it onto the WKB solutions.  Doing so will show that no new corrections to the scattering coefficients are necessary.  Finally, we compare the new local solutions to numerical solutions in order to show the improved matching.

\subsection{Effects of Quadratic terms on the WKB solutions\label{sec:coupled_WKB}}


\begin{figure}
\begin{center}
\includegraphics[scale=0.6]{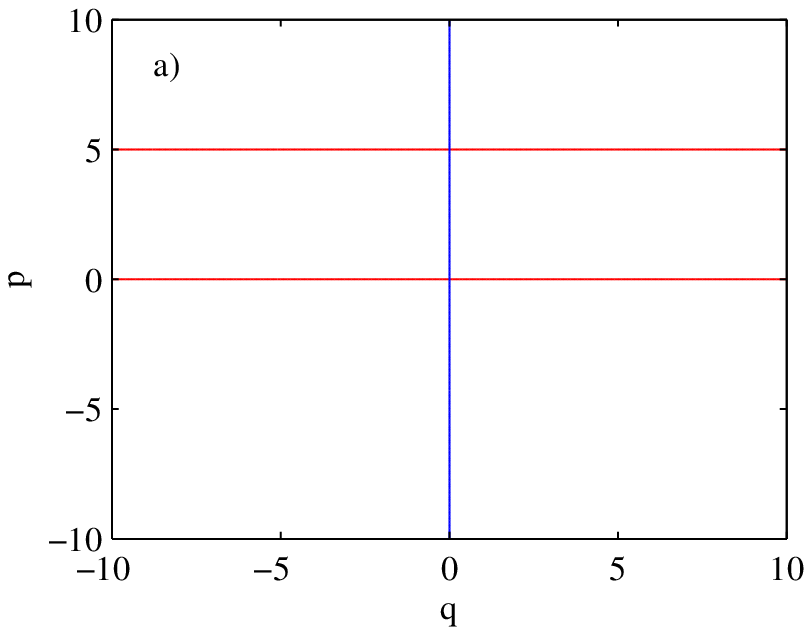}
\includegraphics[scale=0.6]{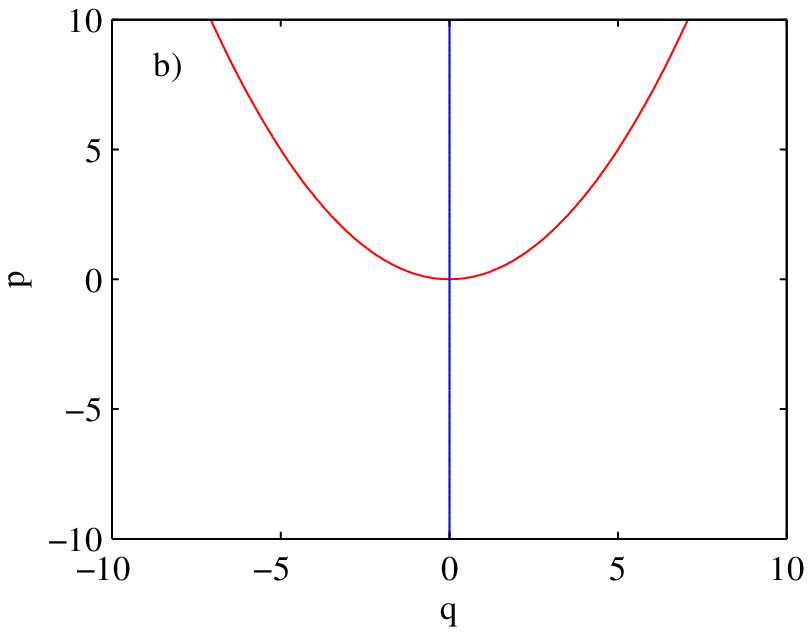}\\
\includegraphics[scale=0.6]{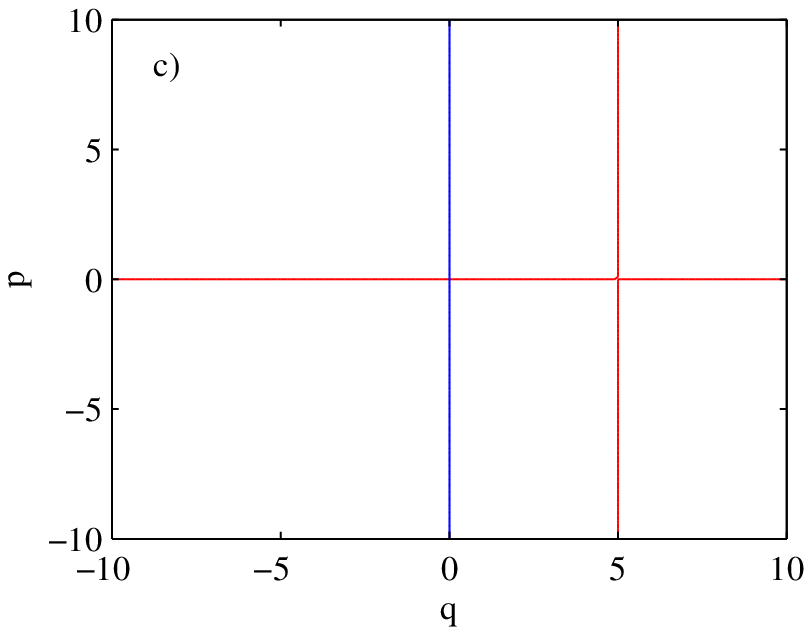}
\includegraphics[scale=0.6]{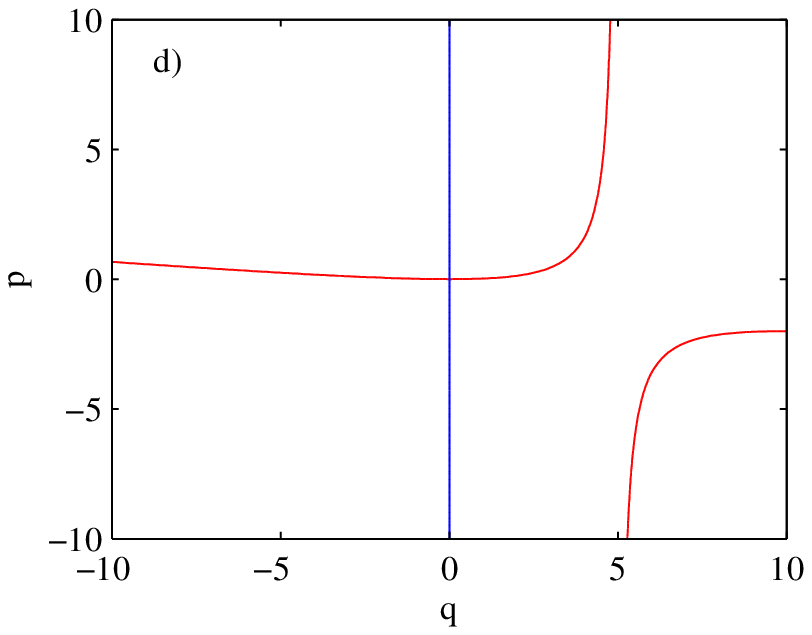}
\end{center}
\caption{\label{fig:dispersion_higher}
Dispersion curves for the uncoupled modes, showing the effect of the higher order terms ($\epsilon = 1/5$).  The red curves are solutions to $D_{11}=0$, and the blue curves are solutions to $D_{22}=0$.  The original crossing is at $q=p=0$.  For all plots $a_2=b_2=c_2=0$.  The remaining parameter values used for these plots are: (a) $a_1 = 1, \, b_1=c_1=0$:  This term introduces a new branch to the dispersion surface $D_{11}=0$ at $p=1/a_1\epsilon$.  As $\epsilon\rightarrow 0$, this branch moves off to infinity.   (b) $c_1 = 1,  \, a_1=b_1=0$:  This term introduces a curvature into the $D_{11}$ dispersion surface.  In the $p$ representation, an uncoupled solution in the upper channel would now look like an Airy function.  (c) $b_1 = 1, \, a_1=c_1=0$:  Although not obvious from the dispersion curves, this term introduces an amplitude variation which has a square-root form, and which matches the WKB amplitude variation due to action conservation.  This term also introduces a resonance in the upper channel at $q=1/b_1\epsilon$.  (d) $b_1 = 1$, $c_1 = 0.1, \, a_1=0$:  In general, there will be several higher order terms, all of which effect the dispersion surfaces, leading to phase shifts, amplitude variations, and resonances.
}
\end{figure}


The calculation of the WKB solution proceeds as in Section \ref{sec:coeffs}, where the WKB solution was calculated to linear order in the phase space variables.  Introduction of the higher order terms will modify the eigenvalues of the dispersion matrix, and introduce corrections of $\mathcal{O}(\epsilon)$ into the WKB solutions.

First, we can use the equation $\det \mathbf{D}(q,p) = 0$ to find the WKB phase, since the determinant is only zero when one of the eigenvalues is zero.  When we include the second order terms in the dispersion matrix, this equation becomes  
\begin{equation}\label{eq:quad_det}
\det \mathbf{D}(q,p) = (-p + \epsilon \mathcal{D}_1(q,p))(q+\epsilon \mathcal{D}_2(q,p)) -|\tilde\eta|^2 =0,
\end{equation}
where the second order terms are 
\begin{equation}
\mathcal{D}_1(q,p) = a_1 p^2 +b_1qp + c_1 q^2, \qquad
\mathcal{D}_2(q,p) = a_2 q^2 +b_2qp + c_2 p^2.
\end{equation}
Expand the solution $p(q)$ in powers of $\epsilon$
\begin{equation}
p(q) = p_0(q) + \epsilon p_1(q) + \ldots 
\end{equation}
and insert this into Equation (\ref{eq:quad_det}).  The terms of order $\epsilon^0$ give
\begin{equation}
-pq -|\tilde\eta|^2=0,
\end{equation}
and so the first term in $p(q)$ is 
\begin{equation}
p_0(q) = \frac{-|\tilde\eta|^2}{q}.
\end{equation}
The order $\epsilon$ terms in Equation (\ref{eq:quad_det}) can be solved for $p_1(q)$:
\begin{eqnarray}
p_1(q) &=& \frac{1}{q} \big( -p_0 \mathcal{D}_2(q,p_0) + q\mathcal{D}_1(q,p_0) \big) \\
&=& \frac{|\tilde\eta|^2}{q^2} \mathcal{D}_2(q,p_0) + \mathcal{D}_1(q,p_0) \\
&=& c_1 q^2 + |\tilde\eta|^2 (a_2 -b_1) 
+ \frac{|\tilde\eta|^4}{q^2} (a_1 -b_2) +\frac{c_2 |\tilde\eta|^6}{q^4}.
\end{eqnarray}
The term $c_1 q^2$ comes from the curvature of the uncoupled dispersion surface for the upper channel.  The $\mathcal{O}(|\tilde\eta|^2)$ term is the leading order effect of the coupling.  It enters the WKB solution at order $\epsilon |\tilde\eta|^2$.  The terms with negative powers of $q$ only become large for small $q$, but they cause the WKB phase to diverge in the mode conversion region:
\begin{eqnarray} 
 \psi_1(q) &= A_1(q) \exp \left( i \int^q p_0(q') + \epsilon p_1(q') \,dq'  \right) \\
 &= A_1(q) \exp \left\{-i|\tilde\eta|^2 \ln (q) \right. \nonumber\\
 &\qquad \left. +i\epsilon\left(\frac{c_1 q^3}{3} 
+ |\tilde\eta|^2 (a_2 -b_1) q 
- \frac{|\tilde\eta|^4}{q} (a_1 -b_2) -\frac{c_2 |\tilde\eta|^6}{3 q^3}
\right)\right\}. \label{eq:wkb_coupled_1}
\end{eqnarray}

The WKB amplitude is computed from the derivative of the eigenvalue $\lambda_\alpha$ which approaches $D_{11}$ for $|q|\gg 1$, and which is the generator for ray evolution in the upper channel:
\begin{eqnarray}
\lambda_\alpha = \frac{D_{11}+D_{22}}{2} - {\rm sgn}(q) \sqrt{\left(\frac{D_{22}-D_{11}}{2} \right)^2 +|\tilde\eta|^2} \,.
\end{eqnarray}
The $p$ derivative of this has order $\epsilon$ terms because of the order $\epsilon$ terms in $D_{11}$ and $D_{22}$, and also because of the order $\epsilon$ term in $p(q)$:
\begin{eqnarray}
\fl \partial_p \lambda_\alpha|_{p(q)} &= \left.\left\{-\frac{1}{2} -\frac{{\rm sgn}(q)}{2} \left(\frac{q+p}{\sqrt{(p+q)^2+4|\tilde\eta|^2}}  \right)  \right\}\right|_{p_0(q)+\epsilon p_1(q)} \nonumber\\
\fl &\quad + \epsilon \left\{  
\frac{1}{2} (\partial_p\mathcal{D}_1 + \partial_p\mathcal{D}_2) 
-\frac{{\rm sgn}(q)}{2 \sqrt{\left(q+p \right)^2 +4|\tilde\eta|^2}}
\left.\bigg[(q+p)(\partial_p\mathcal{D}_2 -\partial_p\mathcal{D}_1)\right.\right. \nonumber\\
\fl &\quad \left.\left.\left.+(\mathcal{D}_2 -\mathcal{D}_1)  
-\frac{ (q+p)^2 (\mathcal{D}_2 -\mathcal{D}_1)}{\left(q+p \right)^2 +4|\tilde\eta|^2}
 \right]  \right\} \right|_{p_0(q)} + \mathcal{O}(\epsilon^2).
\end{eqnarray}
As a first step in evaluating this derivative, we can find the leading order effect in $\epsilon$ of the $p_0(q) + \epsilon p_1(q)$ evaluation:
\begin{eqnarray}
\fl \Bigg\{-\frac{1}{2} -\frac{{\rm sgn}(q)}{2} \left(\frac{q+p}{\sqrt{(p+q)^2+4|\tilde\eta|^2}}  \right)  \left.\Bigg\}\right|_{p_0(q)+\epsilon p_1(q)} \nonumber\\
\fl \qquad =\left\{-\frac{1}{2} -\frac{{\rm sgn}(q)}{2} 
\left(\frac{q+p_0 + \epsilon p_1}{\sqrt{(p_0+q)^2+4|\tilde\eta|^2}}  
- \frac{\epsilon p_1(q+p_0)^2}{[(p_0+q)^2+4|\tilde\eta|^2]^{3/2}}\right)  \right\}
+\mathcal{O}(\epsilon^2).
\end{eqnarray}
This gives the expansion of $\partial_p \lambda_\alpha$ in $\epsilon$.  Further simplification can be achieved by considering the denominators:
\begin{eqnarray}
[(q+p_0)^2 + 4|\tilde\eta|^2]^{-\kappa/2} & = \left[\left(q-\frac{|\tilde\eta|^2}{q}\right)^2 + 4|\tilde\eta|^2\right]^{-\kappa/2} \\
&= \left[q^2-2q\frac{|\tilde\eta|^2}{q}+\left(\frac{|\tilde\eta|^2}{q}\right)^2 + 4|\tilde\eta|^2\right]^{-\kappa/2} \\
&=\left[\left(q+\frac{|\tilde\eta|^2}{q}\right)^2\right]^{-\kappa/2}  \\
&=|q|^{-\kappa} \Bigg(1-\kappa \frac{|\tilde\eta|^2}{q^2}
+\frac{\kappa}{2} (\kappa + 1)\frac{|\tilde\eta|^4}{q^4} \nonumber\\
&\qquad -\frac{\kappa}{6}(\kappa + 1)(\kappa + 2)\frac{|\tilde\eta|^6}{q^6} +\ldots\Bigg).
\end{eqnarray}
Since we are working with the WKB approximation, $q\gg |\tilde\eta|^2$, and we can truncate this expansion.  The order at which it can be truncated will be determined by the highest power of $q$ by which this term is multiplied.  For example, when multiplied by $p_1 q^2$, which is $\mathcal{O}(q^4)$, we need to keep terms through $\mathcal{O}(q^{-6})$.
\mycomment{
We can also make the substitution
\begin{eqnarray}
q+p_0 = q \left(1-\frac{|\tilde\eta|^2}{q^2}\right).
\end{eqnarray}
}

\mycomment{ 
\begin{eqnarray}
\partial_p \lambda_\alpha|_{p(q)} &= \left(-\frac{1}{2}-\frac{1}{2}\left(1-\frac{|\tilde\eta|^2}{q^2}\right)^2 \right) \nonumber\\
&\quad -\frac{\epsilon}{2} {\rm sgn}(q)p_1 (q+p_0)^2\left(1-3\frac{|\tilde\eta|^2}{q^2}\right) |q|^{-3} \nonumber\\
&\quad + \epsilon \left\{  
\frac{1}{2} (\partial_p\mathcal{D}_1 + \partial_p\mathcal{D}_2) 
-{\rm sgn}(q)|q|^{-1}\left(1-\frac{|\tilde\eta|^2}{q^2}\right)
\left.\bigg[(q+p_0)(\partial_p\mathcal{D}_2 -\partial_p\mathcal{D}_1)\right.\right. \nonumber\\
&\quad \left.\left.+p_0(\mathcal{D}_2 -\mathcal{D}_1)  
+  \frac{p_0 (q+p_0)^2 (\mathcal{D}_2 -\mathcal{D}_1)}{2}
 \right]  \right\} .
\end{eqnarray}
By treating $1 / q$ as a small parameter, this can be further simplified, giving
\begin{eqnarray}
\partial_p \lambda_\alpha|_{p(q)} &= -1+\frac{|\tilde\eta|^2}{q^2}
-\frac{\epsilon}{2} \left(1-\frac{5|\tilde\eta|^2}{q^2}\right)\frac{p_1}{q}
+ \frac{\epsilon}{2} (\partial_p\mathcal{D}_1+\partial_p\mathcal{D}_2)\nonumber\\
&\quad -\epsilon\left(1-\frac{2|\tilde\eta|^2}{q^2}\right) (\partial_p\mathcal{D}_2-\partial_p\mathcal{D}_1)
+\epsilon\frac{|\tilde\eta|^2}{q^2}(\mathcal{D}_2 -\mathcal{D}_1) \left(1+\frac{q^2}{2} - \frac{|\tilde\eta|^2}{2}\right).
\end{eqnarray}
}

Inserting all of this, including the expressions for $p_1$, $\mathcal{D}_1$, and $\mathcal{D}_2$, back into the derivative, and gathering terms of like powers in $q$ gives
\begin{eqnarray}
\fl \partial_p \lambda_\alpha|_{p(q)} &= 
-1  + \frac{|\tilde\eta|^2}{q^2} + \epsilon b_1 q + \epsilon |\tilde\eta|^2 \left(b_2 - b_1 -2a_2 -2 a_1 \right) q^{-1} + \mathcal{O}(\epsilon^2, q^{-3}).
\end{eqnarray}
The first three terms are the most significant in the limit we are considering.  So, the WKB amplitude can be written
\begin{equation}
A_1(q) = \left| \frac{\partial \lambda_{\alpha} (q,p)}{\partial p} \right|_{p=p(q)}^{-1/2} \approx
 \left|-1+\frac{|\tilde\eta|^2}{q^2} +\epsilon b_1 q\right|^{-1/2}.
\end{equation}

The calculation for the coupled WKB mode in the lower channel proceeds along similar lines, and uses the $p$ representation of the equations.  The result is
\begin{equation}
\fl \psi_2(p) = A_2(p) \exp \left\{i|\tilde\eta|^2 \ln (p) +i\epsilon\left(\frac{c_2 p^3}{3} + |\tilde\eta|^2 (a_1 -b_2) p 
- \frac{|\tilde\eta|^4}{p} (a_2 -b_1) -\frac{c_1 |\tilde\eta|^6}{3 p^3}
\right)\right\}, \label{eq:wkb_coupled_2}
\end{equation}
with
\begin{equation}
A_2(p) = \left|1+\epsilon b_2 p - \frac{|\tilde\eta|^2}{p^2} + \ldots \right|^{-1/2}.
\end{equation}

We will use these expressions for the coupled WKB solutions when matching the local solution to the propagating modes.

\subsection{Local Coupled Solutions \label{sec:local_sol}}

Now that we know the form of the incoming and outgoing WKB waves, we need to solve the system of equations locally in order to find the scattering coefficients which connect the incoming to outgoing waves.  
We will find the higher order corrections by expanding the local fields in $\epsilon$, and then using the $\mathcal{O}(\epsilon)$ equations to get the local solution.  Write the fields as asymptotic expansions in $\epsilon$:
\begin{equation} \label{eq:psi1_q}
\psi_1(q) = q^{-i|\tilde\eta|^2} (1 + \epsilon \Theta_1(q) + \mathcal{O}(\epsilon^2) ),
\end{equation}
\begin{equation} \label{eq:psi2_q}
\psi_2(q) = -\tilde\eta^* q^{-i|\tilde\eta|^2-1} (1 + \epsilon \Theta_2(q) + \mathcal{O}(\epsilon^2) ) .
\end{equation}
Because of the form of the equations, it is convenient to expand $\Theta_1$ and $\Theta_2$ as power series in $q$:
\begin{equation}
\Theta_1(q) = \sum_{n=-\infty}^\infty s_n q^n, \quad \Theta_2(q) = \sum_{n=-\infty}^\infty \tilde s_n q^n.
\end{equation}
The coefficients $s_n$ and $\tilde s_n$ are constants which depend on $\epsilon$.


In order to compact the notation, and simplify the calculations, define $\alpha_n$ as the coefficient obtained when acting with $\hat p$ on one of the terms in the series expansion above:
\begin{eqnarray}
	\hat p \, q^{-i|\tilde\eta|^2 + n} &= -i \partial_q q^{-i|\tilde\eta|^2 + n} \\
	& = i(i|\tilde\eta|^2 -n) q^{-i|\tilde\eta|^2 + n - 1} \\
	& \equiv \alpha_n q^{-i|\tilde\eta|^2 + n - 1}.  \label{eq:def_alpha}
\end{eqnarray}

Our equations involve the operators $\hat p$, $\hat p^2$, $\hat q^2$, and $\widehat{pq}$ \footnote{By $\widehat{pq}$ we mean the operator whose Weyl symbol is $pq$.  Since $p$ and $q$ commute as variables in phase space, we could also have written this operator as $\widehat{qp}$}.  Using the definition of $\alpha_n$ above, we can evaluate the action of these operators for an arbitrary term in our series:  
\begin{eqnarray}
\hat p \, q^{-i|\tilde\eta|^2 + n} &= \alpha_n q^{-i|\tilde\eta|^2 + n - 1} .
\end{eqnarray}
Applying this formula twice gives us
\begin{equation}
\hat p^2 \, q^{-i|\tilde\eta|^2 + n} = \alpha_n \alpha_{n-1} q^{-i|\tilde\eta|^2 +n -2}.
\end{equation}
Similarly we can calculate the action of the symmetrized operator $\widehat{pq}$:
\begin{eqnarray}
\widehat{pq} \, q^{-i|\tilde\eta|^2 + n} &=  \frac{1}{2} (\hat p \hat q + \hat q \hat p)  q^{-i|\tilde\eta|^2 + n} \\
&= \frac{1}{2} (2 \hat q \hat p - i)  q^{-i|\tilde\eta|^2 + n} \\
&= \left( \alpha_n -\frac{i}{2} \right) q^{-i|\tilde\eta|^2 + n}.
\end{eqnarray}
We will also use
\begin{equation}
\hat q^2 \, q^{-i|\tilde\eta|^2 + n} = q^{-i|\tilde\eta|^2 + n+2}.
\end{equation}

We can now write out our system of equations in (\ref{eq:2nd_order_wave_eqn}) using the coefficients $\alpha_n$ and our series expansions for the fields.  Keeping only terms of order $\epsilon$, we get
\begin{equation}
\fl \hat D_{11}\psi_1(q) + \tilde\eta \psi_2(q)  =-\hat{p} \sum_n s_n q^{-i|\tilde\eta|^2 + n} + \mathcal{D}_1(\hat q, \hat p) q^{-i|\tilde\eta|^2} - |\tilde\eta|^2\sum_n \tilde s_n q^{-i|\tilde\eta|^2 -1 + n} =0
\end{equation}
and
\begin{equation}
\fl \tilde\eta^* \psi_1(q) + \hat D_{22} \psi_2(q) = 
\sum_n s_n q^{-i|\tilde\eta|^2 + n} - \mathcal{D}_2(\hat q, \hat p) q^{-i|\tilde\eta|^2-1} - q\sum_n \tilde s_n q^{-i|\tilde\eta|^2 -1 + n} =0 .
\end{equation}
Now evaluate the operators in these equations using the expressions derived above:
\begin{eqnarray}
\sum_n \left( -\alpha_n s_n - |\tilde\eta|^2 \tilde s_n \right) q^{-i|\tilde\eta|^2 -1 + n} \nonumber\\
+ a_1 \left( \alpha_0 \alpha_{-1} \right) q^{-i|\tilde\eta|^2 -2} 
+ b_1\left( \alpha_{0} -\frac{i}{2}  \right) q^{-i|\tilde\eta|^2}
+ c_1 q^{-i|\tilde\eta|^2 +2} =0,
\end{eqnarray}
\begin{eqnarray}
\sum_n \left( s_n - \tilde s_n \right) q^{-i|\tilde\eta|^2 + n} \nonumber\\
- c_2 \left(\alpha_{-1}\alpha_{-2} \right) q^{-i|\tilde\eta|^2 -3} 
 - b_2\left( \alpha_{-1} -\frac{i}{2} \right) q^{-i|\tilde\eta|^2-1}
- a_2 q^{-i|\tilde\eta|^2 +1} =0.
\end{eqnarray}
These equations can now be solved for the coefficients $s_n$ and $\tilde s_n$, by matching like powers of $q$.  For $n \not\in \{-3,-1,1,3 \}$ these give $s_n = \tilde s_n =0$.  Otherwise, we have a set of linear equations which can be solved (using the definition of $\alpha_n$ to simplify the coefficients) to give
\begin{eqnarray}
s_3 &=&  \frac{ic_1}{3} \\
\tilde s_3 &=& \frac{ic_1}{3} \\
s_1&=&  i(b_1-a_2)\alpha_0 +\frac{b_1}{2} \\
\tilde s_1 &=&  i(b_1-a_2)\alpha_1 -\frac{b_1}{2} \\
s_{-1} &=&  i(b_2-a_1)\alpha_0\alpha_{-1} +\frac{b_2}{2}  \alpha_0 \\
\tilde s_{-1} &=& i(b_2-a_1) \alpha_0\alpha_{-1} -\frac{b_2}{2} \alpha_{-1} \\
s_{-3} &=& \frac{ic_2}{3} \alpha_0\alpha_{-1}\alpha_{-2} \\
\tilde s_{-3} &=& \frac{i c_2}{3} \alpha_{-1}\alpha_{-2}\alpha_{-3}.
\end{eqnarray}

These coefficients give the corrected local solutions in the $q$ representation, which can be matched to the WKB solutions for the upper channel.  For the lower channel, we need the local solution in the $p$ representation.  The calculation for the transformation to the $p$ representation involves integrating functions which have singularities at the origin, just as we had when calculating the conversion coefficient in Equation (\ref{eq:con_coeff});
\begin{eqnarray}
\psi_2(p) &= \frac{1}{\sqrt{2\pi i}}\int dq\, e^{-ipq} \psi_2(q) \\
&= \frac{1}{\sqrt{2\pi i}}\int dq\, e^{-ipq} (-\tilde\eta^* q^{-i|\tilde\eta|^2 -1}) ( 1+ \epsilon \Theta_2(q) +\mathcal{O}(\epsilon^2) ) \\
&=\frac{1}{\sqrt{2\pi i}}\int dq\, e^{-ipq} (-\tilde\eta^* q^{-i|\tilde\eta|^2 -1})( 1+ \epsilon \sum_n \tilde s_n q^n +\mathcal{O}(\epsilon^2) ).
\end{eqnarray}
This integral can be evaluated (by putting it into the form of Hankel's contour integral for the complex Gamma function), and the answer written as the power series
\begin{equation}\label{eq:psi2_p}
\psi_2(p) = \frac{\beta^*}{\tau} p^{i|\tilde\eta|^2} \left( 1+ \epsilon \sum_{n=-\infty}^\infty \tilde\sigma_n p^n+ \mathcal{O}(\epsilon^2) \right),
\end{equation}
where terms in the series can be shown to be
\begin{equation}
\tilde\sigma_{n} = \left\{\begin{array}{cc}
\alpha_0 \alpha_1 \cdots \alpha_{-n-1} \tilde s_{-n} , & n<0 \\
0, & n=0 \\
\frac{\tilde s_{-n}}{\alpha_{-1}\alpha_{-2}\cdots \alpha_{-n}}, & n>0.
\end{array}
\right.
\end{equation}
Inserting the values of $\tilde s_n$ from above gives
\begin{eqnarray}
\tilde \sigma_{-3} &= \frac{ic_1}{3} \alpha_2 \alpha_1 \alpha_0 \\
\tilde \sigma_{-1} &=  i(b_1-a_2)\alpha_1 \alpha_0 -\frac{b_1}{2} \alpha_0 \\
\tilde \sigma_{1} &= i(b_2-a_1) \alpha_0 -\frac{b_2}{2}  \\
\tilde \sigma_{3} &= \frac{i c_2}{3} .
\end{eqnarray}
\mycomment{
\begin{eqnarray}
\sigma_{-3} &=&  \frac{ic_1}{3} \frac{\beta_{3}}{\beta_0} \\
\tilde \sigma_{-3} &=& \frac{ic_1}{3} \frac{\beta_{2}}{\beta_{-1}}\\
\sigma_{-1}&=&  i(b_1-a_2)\frac{\beta_1}{\beta_{-1}} +\frac{b_1}{2} \frac{\beta_{1}}{\beta_0}\\
\tilde \sigma_{-1} &=&  i(b_1-a_2)\frac{\beta_1}{\beta_{-1}} -\frac{b_1}{2} \frac{\beta_0}{\beta_{-1}} \\
\sigma_{1} &=&  i(b_2-a_1) \frac{\beta_{-1}}{\beta_{-2}} +\frac{b_2}{2}   \\
\tilde \sigma_{1} &=& i(b_2-a_1) \frac{\beta_0}{\beta_{-1}} -\frac{b_2}{2}  \\
\sigma_{3} &=& \frac{ic_2}{3} \\
\tilde \sigma_{3} &=& \frac{i c_2}{3} .
\end{eqnarray}

Inserting these into the expression for $\phi_2(p)$ gives
\begin{eqnarray}
	\phi_2(p) = \frac{\beta^*}{\tau} p^{i|\tilde\eta|^2} \left( 1+ \epsilon( \frac{ic_1}{3} \alpha_2 \alpha_1 \alpha_0 p^{-3} + (i(b_1-a_2)\alpha_1 \alpha_0 -\frac{b_1}{2} \alpha_0) p^{-1} + (i(b_2-a_1) \alpha_0 -\frac{b_2}{2}) p + \frac{i c_2}{3} p^3 ) + \mathcal{O}(\epsilon^2) \right)
\end{eqnarray}
}
These coefficients, together with the coefficients $s_n$ above, give the new local solutions, which include the effects of the quadratic terms.

\subsection{Matching the Local and Far-field Solutions\label{sec:2nd_matching}}

\subsubsection{Matching the Upper Channel}

In order to find the connection between the incoming and outgoing wave fields, we need to match the WKB solutions in Section \ref{sec:coupled_WKB} with the local solutions in Section \ref{sec:local_sol}.  This matching will allow us to find the scattering matrix for this mode conversion, which determines the amplitudes and phases of the outgoing waves given the amplitudes and phases of the incoming waves.  The various WKB waves can be identified with the dispersion surfaces from which their phases are calculated.  This lets us label the branches of the dispersion surface by the type of WKB mode which it generates, as in Figure \ref{fig:uncoupled_modes}.  

Matching a WKB wave incoming in the upper channel requires comparison of the WKB and local solutions for the upper channel.  The WKB solution is given in Equation (\ref{eq:wkb_coupled_1}) and the local solution is written as the series in Equation (\ref{eq:psi1_q}); 
\begin{eqnarray}\label{eq:wkb_expansion_1}
\fl \psi_1^{({\rm WKB})}(q) 
&= q^{-i|\tilde\eta|^2}\left( 1 + \frac{i\epsilon c_1 q^3}{3} + \frac{\epsilon b_1 q}{2} + i\epsilon (a_2-b_1)|\tilde\eta|^2 q +\mathcal{O}(\epsilon^2,q^{-1}) \right)
\end{eqnarray} 
\mycomment{&= \frac{q^{-i| \tilde\eta|^2} e^{i\epsilon (c_1 q^3/3 +(a_2 -b_1)| \tilde\eta|^2 q)}  }{\sqrt{1-\epsilon b_1 q +| \tilde\eta|^2/q^2}} \\}
and
\begin{eqnarray}
\fl \psi_1^{({\rm Local})} (q) &= q^{-i|\tilde\eta|^2}  \left(1+   
\frac{i\epsilon c_1 q^3}{3} + \frac{\epsilon b_1 q}{2} + i\epsilon (a_2-b_1)|\tilde\eta|^2 q
+\mathcal{O}(\epsilon^2,q^{-1})  \right) .
\end{eqnarray}
Here, the terms with negative powers of $q$ have been dropped, since the effect of these terms is localized to the mode conversion region.  

The incoming  matching point $q_{MI} < 0$ is now chosen where both of these expressions are valid, and where the neglected inverse powers of $q$ are actually negligible.  The amplitude and phase of the incoming WKB mode at this point are used to set the incoming amplitude and phase of the local solution;
\begin{eqnarray}
\psi_1'(q) &=& A \frac{ \psi_1^{({\rm WKB})}(q_{MI})  }{\psi_1^{({\rm Local})} (q_{MI}) } \psi_1^{({\rm Local})} (q) \\
&=& A \psi_1^{({\rm Local})} (q) .
\end{eqnarray}
Here $A$ is the complex amplitude of the incoming wave.  As in the case of the linear order problem in Section \ref{sec:lin}, we must choose a branch cut for the term $e^{-i|\tilde\eta|^2 \ln q}$.  This choice will let us write the matched local solution in terms of the magnitude of $q$, 
\begin{equation} \label{eq:phi1_prime}
\fl \psi_1'(q) = \cases{
A e^{\pi |\tilde\eta|^2} |q|^{-i|\tilde\eta|^2}\left(1+ i\epsilon |\tilde\eta|^2 (a_2-b_1)q + \frac{\epsilon b_1 q}{2} + \frac{i \epsilon c_1 q^3}{3}
+\mathcal{O}(\epsilon^2)  \right) &for $q<0$\\
A |q|^{-i|\tilde\eta|^2}\left(1+ i\epsilon |\tilde\eta|^2 (a_2-b_1)q + \frac{\epsilon b_1 q}{2} + \frac{i \epsilon c_1 q^3}{3}
+\mathcal{O}(\epsilon^2)  \right) &for $q>0$
}.
\end{equation}

Now choose a matching point $q_{MO} > 0$ where this local solution can be matched to the outgoing WKB wave.  We then can write the outgoing field as
\begin{eqnarray}
\fl \psi_1''(q) &= \frac{\psi_1'(q_{MO})}{\psi_1^{({\rm WKB})}(q_{MO})}\psi_1^{({\rm WKB})}(q) \\
\fl &= A |q|^{-i|\tilde\eta|^2}\left(1+ i\epsilon |\tilde\eta|^2 (a_2-b_1)q + \frac{\epsilon b_1 q}{2} + \frac{i \epsilon c_1 q^3}{3} +\mathcal{O}(\epsilon^2,q^{-2})  \right)
\end{eqnarray}
This expression shows that the transmission coefficient $\tau$ remains the same as it was in the case of the linearized problem described in Section \ref{sec:lin}.  The quadratic order terms introduced corrections to the local solution which makes the local solution better match the WKB solutions.

\subsubsection{Matching the Lower Channel}

Use the $p$ representation to match the lower channel.  Match the WKB solution from Equation (\ref{eq:wkb_coupled_2}) ,
\begin{eqnarray}
\fl \psi_2^{({\rm WKB})}(p) 
&= p^{i|\tilde\eta|^2} \left( 1 +  i\epsilon |\tilde\eta|^2 (a_1-b_2)p - \frac{\epsilon b_2 p}{2} + \frac{i\epsilon c_2 p^3}{3} + \mathcal{O}(\epsilon^2,p^{-1}) \right)
\end{eqnarray}
\mycomment{&= \frac{p^{i| \tilde\eta|^2} e^{i\epsilon c_2 p^3/3 +  i\epsilon | \tilde\eta|^2 (a_1-b_2)p}  }{\sqrt{1+\epsilon b_2 p + | \tilde\eta|^2/p^2}} \\}
to the local solution from the series given in  Equation (\ref{eq:psi2_p}),
\begin{equation}
\fl \psi_2^{({\rm Local})} (p) = \frac{\beta^*}{\tau} p^{i|\tilde\eta|^2} \left(  1+ 
 i\epsilon |\tilde\eta|^2 (a_1-b_2)p - \frac{\epsilon b_2 p}{2} + \frac{i \epsilon c_2 p^3}{3}
 + \mathcal{O}(\epsilon^2,p^{-1})
 \right),
\end{equation}
where, as for the upper channel, we have dropped the terms which involve inverse powers of $p$, since they are negligible outside of the mode conversion region.
The matching proceeds as in the first channel, except that the initial amplitude and phase are given by the incoming data in the upper channel.  Since the local solutions given in Equations (\ref{eq:psi1_q}) and (\ref{eq:psi2_q}) have the correct relative amplitude and phase, we can use the overall initial amplitude and phase from (\ref{eq:phi1_prime}) as the initial amplitude and phase for the lower channel.  This gives us the local field for the lower channel;
\begin{equation}
\psi_2'(p) = A \tau \psi_2^{({\rm Local})}(p).
\end{equation}
We now pick a point $p_M>0$ where we match the local solution onto the outgoing WKB wave.  The outgoing wave is then
\begin{eqnarray}
\fl \psi_2''(p) &= A \tau  \frac{\psi_2^{({\rm Local})}(p_M)}{\psi_2^{({\rm WKB})}(p_M)} \psi_2^{({\rm WKB})}(p) \\
\fl &= A \beta^*p^{i|\tilde\eta|^2}\left( 1 +  i\epsilon |\tilde\eta|^2 (a_1-b_2)p - \frac{\epsilon b_2 p}{2} + \frac{i\epsilon c_2 p^3}{3} + \mathcal{O}(\epsilon^2,p^{-1}) \right).
\end{eqnarray}
This shows that the conversion coefficient is unchanged, and that the local solution now matches the WKB solution much better.

\section{Comparison with Numerical Solution}

In order to demonstrate that the corrected solutions derived above do actually match better, we compared both the corrected local fields and the matched WKB waves to the results of numerical simulations.  In order to avoid the singularities in the solutions, the numerical calculation was carried out in the $Q$ representation, where the phase space coordinates $(Q,P)$ are those defined by the linear canonical transformation in Equation (\ref{eq:Q_rep_transformation}).
The associated metaplectic transformations of the coupled WKB waves from Equations (\ref{eq:wkb_coupled_1}) and (\ref{eq:wkb_coupled_2}) are [without the divergent inverse powers in the phase]
\begin{equation}\label{eq:met_1}
\psi_1^{({\rm WKB})}(Q) = \int_{-\infty}^\infty e^{iF_1(Q,q)} \frac{q^{-i| \tilde\eta|^2} e^{i\epsilon c_1 q^3/3 +i\epsilon | \tilde\eta|^2 (a_2 -b_1) q}  }{\sqrt{1-\epsilon b_1 q -| \tilde\eta|^2/q^2}} \, dq ,\quad |q|>1
\end{equation}
and
\begin{equation}\label{eq:met_2}
\psi_2^{({\rm WKB})}(Q) = \int_{-\infty}^\infty e^{iF_2(Q,p)} \frac{p^{i| \tilde\eta|^2} e^{i\epsilon c_2 p^3/3 +  i\epsilon | \tilde\eta|^2 (a_1-b_2)p}  }{\sqrt{1+\epsilon b_2 p - | \tilde\eta|^2/p^2}} \, dp ,\quad |p|>1, 
\end{equation}
where $F_1(Q,q)$ and $F_2(Q,p)$ are the generating functions for the linear canonical transformation:
\begin{eqnarray}
F_1(Q,q) &= \frac{1}{2} (Q^2 - 2\sqrt{2} Qq + q^2) ,\\
F_2(Q,p) &= -\frac{1}{2} (Q^2 - 2\sqrt{2} Qp + p^2) .
\end{eqnarray}
The conjugate variables are given by derivatives of the generating functions:
\begin{eqnarray}
P = \frac{\partial F_1}{\partial Q}, &\qquad  p = - \frac{\partial F_1}{\partial q},  \label{eq:gen_derivs_1} \\ 
P = \frac{\partial F_2}{\partial Q}, &\qquad  q = \frac{\partial F_2}{\partial p} .
\end{eqnarray}

In order to evaluate the metaplectic integrals in equations (\ref{eq:met_1}) and (\ref{eq:met_2}), we write $q$ in terms of $Q$ and possibly also derivatives with respect to $Q$, i.e.\ we find the $Q$ representation of the operator $\hat q$.  The properties of the generating functions allow us to do this.  First, use Equations (\ref{eq:Q_rep_transformation}) and (\ref{eq:gen_derivs_1}) to write
\begin{equation}
q = \frac{1}{\sqrt 2} (Q-P)  = \frac{1}{\sqrt 2} (Q-i\partial_Q F_1(Q,q)) .
\end{equation}
Now, notice that we can combine the derivative $\partial_Q F_1(Q,q)$ with the phase in the integral to write
\begin{equation}
\partial_Q F_1(Q,q) e^{i F_1(Q,q)} = -i\partial_Q e^{i F_1(Q,q)}.
\end{equation}
We can now consider the higher order terms in the integral as a psudeodifferential operator acting on the function $q^{-i| \tilde\eta|^2}$, by  making the substitution 
\begin{equation}
q \rightarrow \frac{1}{\sqrt 2} (Q+i \partial_Q ),
\end{equation}
wherever $q$ appears in the higher order terms.  In order to do this, write the higher order terms as a series:
\begin{eqnarray}
S_1(q) & \equiv  \frac{ e^{i\epsilon c_1 q^3/3 +i\epsilon | \tilde\eta|^2 (a_2 -b_1) q}  }{\sqrt{1-\epsilon b_1 q -| \tilde\eta|^2/q^2}} \\
&= \sum_{n=-\infty}^\infty s_n q^n \\
&\rightarrow \sum_{n=-\infty}^\infty \frac{s_n}{\sqrt 2} (Q+i\partial_Q)^n \\
&= S_1\left(\frac{1}{\sqrt 2}  (Q+i\partial_Q) \right) 
\end{eqnarray}

We can now use this to find the $Q$ representation of the field:
\begin{eqnarray}
\psi_1^{({\rm WKB})}(Q) &= \int_{-\infty}^\infty e^{iF_1(Q,q)} S_1(q) q^{-i|\tilde\eta|^2} \, dq \\
&=  S_1\left(\frac{1}{\sqrt 2}  (Q+i\partial_Q) \right) \int_{-\infty}^\infty e^{iF_1(Q,q)} q^{-i|\tilde\eta|^2} \, dq.
\end{eqnarray}
The integral here is the same as the integral in Equation (\ref{eq:psi1_0_Q_int}), and evaluation gives a parabolic cylinder function.  However, we are interested in the WKB limit, where $|q|\gg 1$ and $|Q|\gg 1$.  So, we can evaluate the integral using the stationary phase approximation, which gives:
\begin{eqnarray}
\fl \int_{-\infty}^\infty e^{iF_1(Q,q)} q^{-i|\tilde\eta|^2} \, dq = 
\sqrt{-\frac{\pi}{2}} e^{i\pi/4} e^{-i Q^2/2 -i|\tilde\eta|^2 \ln (\sqrt{2}Q)} + \mathcal{O}(Q^{-2})
\approx \mathcal{N}e^{-i Q^2/2-i|\tilde\eta|^2 \ln (\sqrt{2}Q)} .
\end{eqnarray}

Since the WKB approximation only valid far from the mode conversion, we will drop all terms in $S_1(q)$ with negative powers of $q$.  Also, since the quadratic term in the phase varies more rapidly than the logarithmic term (for the values of $Q$ we are considering), we will consider the action of the derivatives only on the quadratic term.  This gives us
\begin{eqnarray}
\fl &\psi_1^{({\rm WKB})}(Q) \approx  \mathcal{N} \, S_1\!\left(\frac{1}{\sqrt 2}  (Q+i\partial_Q) \right) e^{-i Q^2/2 -i|\tilde\eta|^2 \ln (\sqrt{2}Q)} \\
\fl &\approx \mathcal{N} e^{-i|\tilde\eta|^2 \ln (\sqrt{2}Q)} \, S_1\!\left(\frac{1}{\sqrt 2}  (Q+i\partial_Q) \right) e^{-i Q^2/2} \\
\fl &=\mathcal{N} e^{-i|\tilde\eta|^2 \ln (\sqrt{2}Q)} \left( 1 + \frac{\epsilon}{\sqrt{2}}\left(\frac{b_1}{2}+i | \tilde\eta|^2 (a_2 -b_1)\right) (Q+i\partial_Q) + \frac{i\epsilon c_1}{3\sqrt{2}} (Q+i\partial_Q)^3 \right) e^{-i Q^2/2} \\
\fl &= \mathcal{N} e^{-i|\tilde\eta|^2 \ln (\sqrt{2}Q)} \left( 1 + \left(\frac{\epsilon b_1}{2}+i \epsilon | \tilde\eta|^2 (a_2 -b_1)\right)  (\sqrt{2}Q) + \frac{i\epsilon c_1}{3} \left((\sqrt{2}Q)^3+3i\sqrt{2} Q \right)  \right) e^{-i Q^2/2} \\
\fl &\approx \mathcal{N}e^{-i Q^2/2-i|\tilde\eta|^2 \ln (\sqrt{2}Q)} \,\, \frac{e^{i\epsilon c_1 (\sqrt 2 Q)^3/3 
+ i \epsilon | \tilde\eta|^2 (a_2 -b_1) \sqrt 2 Q
}  }{\sqrt{1+\epsilon (-b_1 + 2 c_1 )\sqrt{2} Q}} \label{eq:psi1_Q_corr}.
\end{eqnarray}
The appearance of the $c_1$ term in the amplitude may be unexpected.  However, under the Fourier transform, the pure phase $\exp(ic_1q^3/3)$ turns into an Airy function, which has amplitude variations.  Converting from the $q$ representation to the $Q$ representation is done with a metaplectic transformation, which is a sort of ``partial'' Fourier transform.  Therefore, it should not be too surprising that the cubic phase in $q$ would give rise to an amplitude variation in $Q$.

A similar analysis can be computed for the lower channel.  Since our corrections are written in the $p$ representation, we need to make the substitution
\begin{equation}
p = \frac{1}{\sqrt 2} (Q+P) \rightarrow \frac{1}{\sqrt 2} (Q-i \partial_Q ).
\end{equation}
Therefore, the WKB mode in the lower channel is
\begin{eqnarray}
\fl &\psi_2^{({\rm WKB})}(Q) \approx  S_2\left(\frac{1}{\sqrt 2}  (Q-i\partial_Q) \right) 
\mathcal{N}_2 e^{i Q^2/2 + i|\tilde\eta|^2 \ln (\sqrt{2}Q)} \\
\fl &\approx \mathcal{N}_2 e^{i|\tilde\eta|^2 \ln (\sqrt{2}Q)} \left( 1 + \frac{\epsilon}{\sqrt{2}}\left(-\frac{b_2}{2}+i|\tilde\eta|^2(a_1-b_2)\right) (Q-i\partial_Q) + \frac{i\epsilon c_2}{3\sqrt{2}} (Q-i\partial_Q)^3  \right) e^{i Q^2/2} \\
\fl &= \mathcal{N}_2 e^{i|\tilde\eta|^2 \ln (\sqrt{2}Q)} \left( 1 + \epsilon\left(-\frac{b_2}{2}+i|\tilde\eta|^2(a_1-b_2)\right) (\sqrt{2}Q) + \frac{i\epsilon c_1}{3} \left((\sqrt{2}Q)^3-3i\sqrt{2} Q \right)  \right) e^{i Q^2/2} \\
\fl &\approx \mathcal{N}_2 e^{i Q^2/2+i|\tilde\eta|^2 \ln (\sqrt{2}Q)} \,\, \frac{e^{i\epsilon c_2 (\sqrt 2 Q)^3/3 + i\epsilon|\tilde\eta|^2(a_1-b_2)\sqrt{2}Q}  }{\sqrt{1+\epsilon (b_2 - 2 c_2 )\sqrt{2} Q}}.\label{eq:psi2_Q_corr}
\end{eqnarray}

We can now compare our analytical expressions (Equations (\ref{eq:psi1_Q_corr}) and (\ref{eq:psi2_Q_corr})) and numerical simulations.  As seen in Figure (\ref{fig:compare_quad}), these corrected solutions correspond closely to the numerical simulations.  The amplitude of the analytical solutions now contain the square root variation which is due to action conservation, so they match the WKB solutions over a much wider range than before, cf. Figure (\ref{fig:compare_linear}).  The phases of the solutions also show good agreement with the numerical simulations.  Notice that the rapid quadratic variation in the phases as been subtracted from these plots, in order to better show the effects of the coupling and the higher order terms.

\section{Conclusions}

In this paper we have shown how to extend the metaplectic formulation of resonant mode conversion to include the effects of quadratic order variations in the dispersion matrix.  A corrected local solution was derived, and matched onto far-field WKB solutions, showing that the transmission and conversion coefficients are unchanged by the new quadratic order terms.  The corrected solution also matches the far-field solutions over a much larger region, as is illustrated by numerical simulations.

\appendix

\section{The Normal Form in One Spatial Dimension\label{sec:nf}}

These three appendices will demonstrate that, for one-dimensional problems, the dispersion matrix can be put into normal form order by order in powers of the phase space variables.  
In the present one-dimensional setting, by `normal form' we mean that the 
off-diagonal terms are constant and that at linear order the new diagonal terms still form a conjugate
pair.  In two, or more, spatial dimensions the normal form is characterized
by the requirement that the diagonals Poisson-commute with the off-diagonals, which ensures that
the off-diagonals are constant following rays generated by the diagonals.  
This will be discussed elsewhere.

In order to put the matrix into normal form, polarization vectors must be chosen so that the off-diagonal elements at higher order vanish.  The strategy for this calculation 
is broken into three parts for clarity.  The first step is to explicitly calculate the transformation which puts the matrix into normal form through second order (\ref{sec:nf}) as a way to introduce key ideas in the simplest setting.  The second step is to outline the calculation which would be needed to put the matrix in normal form to arbitrary order, and show that there are sufficient free parameters in the transformation to achieve normal form (\ref{sec:extension}). 
In both \ref{sec:nf} and \ref{sec:extension} we ignore Moyal corrections,
which significantly complicate the picture.  The final step is to examine the effect of the Moyal corrections (needed when multiplying symbols of operators) which were neglected in the first two steps (\ref{sec:moyal}).

As described in Section \ref{sec:lin}, and shown in \cite{Tracy02,robertandgregannals}, the $2\times 2$ symbol of the dispersion matrix can be put into the following ``normal form'' at linear order:
\begin{equation}
\mathbf{D}_{{\rm NF}}(q,p) = 
\left(
\begin{array}{cc}
-p & \tilde\eta \\
\tilde\eta^* & q
\end{array}
\right).
\end{equation}
Here, $\tilde\eta$ is a constant since we are working in a two-dimensional phase space.  The higher order corrections to this matrix appear at quadratic order in the phase space variables:
\begin{equation}
\mathbf{D}(q,p) = \mathbf{D}_{{\rm NF}}(q,p) + 
\epsilon^2 \mathbf{D}_{2}(q,p)
 + \mathcal{O}(\epsilon^3),
\end{equation}
where $\epsilon$ is a formal parameter introduced to keep track of the ordering.
Each element of $\mathbf{D}_{2}$ can contain terms which are quadratic in the phase space variables $z=(q,p)$.  When needed, we will also write the first order and second order terms by displaying
the monomials:
\begin{equation}
\mathbf{D}_1(q,p)= q\mathbf{D}_{q}+p\mathbf{D}_{p}, \qquad
\mathbf{D}_2(q,p)= q^2\mathbf{D}_{qq}+pq\mathbf{D}_{pq}+p^2\mathbf{D}_{pp},
\end{equation}
where each of these constant $2\times 2$ matrices of coefficients is hermitian and, therefore, 
generically have four real
parameters.  The first order matrices are particularly simple, but the general second order terms
have twelve real parameters.  


Because $\mathbf{D}$ is in normal form to linear order already, we will use a near-identity change 
of polarization basis to bring it into normal form at second order.  We write $\mathbf{Q}$ as
\begin{equation}
\mathbf{Q}(q,p) = \mathbf{1} + \epsilon \mathbf{Q}_1=
\mathbf{1} + \epsilon p\mathbf{Q}_p + \epsilon q\mathbf{Q}_q,
\end{equation}
where $\mathbf{Q}_p$ and $\mathbf{Q}_q$ are constant $2\times 2$ complex
matrices.  Hence, we have eight complex parameters to work with, or, equivalently, sixteen real ones.
Carrying out the congruence $\mathbf{Q}^{\dag}\mathbf{D}\mathbf{Q}$, collecting orders in $\epsilon$,
we find:
\begin{equation}
\fl  
\mathbf{D}'=\mathbf{D}_0+
\epsilon\left[\mathbf{D}_1+\mathbf{Q}^{\dag}_1\mathbf{D}_0+\mathbf{D}_0\mathbf{Q}_1\right]
+
\epsilon^2\left[\mathbf{D}_2+\mathbf{Q}^{\dag}_1\mathbf{D}_1+\mathbf{D}_1\mathbf{Q}_1
+ \mathbf{Q}^{\dag}_1\mathbf{D}_0\mathbf{Q}_1 \right]
+{\mathcal{O}}(\epsilon^3)
\end{equation}
The sixteen real parameters in $\mathbf{Q}$ must be chosen so that the off-diagonals of the
bracketed terms are zero.  The term $\mathbf{D}_1$ is already diagonal, so at ${\mathcal O}(\epsilon)$ we
have the two conditions:
\begin{equation}
\left[\mathbf{Q}^{\dag}_q\mathbf{D}_0+\mathbf{D}_0\mathbf{Q}_q\right]_{12}=0, \qquad
\left[\mathbf{Q}^{\dag}_p\mathbf{D}_0+\mathbf{D}_0\mathbf{Q}_p\right]_{12}=0.
\end{equation}
(The other off-diagonal terms are the complex conjugates of these 
expressions because the congruence preserves the hermiticity.)  Each of these expressions consists of two real conditions on the sixteen parameters in $\mathbf{Q}_1$, hence four of our degrees of freedom are used up.  
These conditions are particularly simple because $\mathbf{D}_0$ is so simple: 
$\left[\mathbf{Q}_q\right]_{11}+\left[\mathbf{Q}^*_q\right]_{22}=0$ and
$\left[\mathbf{Q}_p\right]_{11}+\left[\mathbf{Q}^*_p\right]_{22}=0$.
At ${\mathcal O}(\epsilon^2)$ we have the three conditions:
\begin{equation}
\left[\mathbf{D}_{qq}+\mathbf{Q}^{\dag}_q\mathbf{D}_q+\mathbf{D}_q\mathbf{Q}_q
+ \mathbf{Q}^{\dag}_q\mathbf{D}_0\mathbf{Q}_q
\right]_{12}=0, 
\end{equation}
\begin{equation}
\left[\mathbf{D}_{qp}+\mathbf{Q}^{\dag}_q\mathbf{D}_p+\mathbf{D}_p\mathbf{Q}_q+
\mathbf{Q}^{\dag}_p\mathbf{D}_q+\mathbf{D}_q\mathbf{Q}_p
+ \mathbf{Q}^{\dag}_q\mathbf{D}_0\mathbf{Q}_p + \mathbf{Q}^{\dag}_p\mathbf{D}_0\mathbf{Q}_q
\right]_{12}=0,
\end{equation}
\begin{equation}
\left[\mathbf{D}_{pp}+\mathbf{Q}^{\dag}_p\mathbf{D}_p+\mathbf{D}_p\mathbf{Q}_p
+ \mathbf{Q}^{\dag}_p\mathbf{D}_0\mathbf{Q}_p
\right]_{12}=0, 
\end{equation}
which use up another six parameters. Thus, to ensure that that the off-diagonal
terms are constant to order $\epsilon^3$ we use ten of the sixteen parameters in $\mathbf{Q}_1$.
Notice the pattern: at order $\epsilon^m$ there are $m+1$ monomials $(q^m,q^{m-1}p,\ldots p^m)$.
Hence, at each order we require $m+1$ complex ($2m+2$ real) conditions to be satisfied.
Normalizing the Poisson bracket of the diagonals to linear order so they form a conjugate pair
takes one more parameter.

In summary, by counting parameters, it would appear we can generically
put an arbitrary $2\times 2$ dispersion matrix that is quadratic in the phase space variables into 
normal form.  We now perform an explicit
calculation choosing a particular parameterization to demonstrate this concretely.  To simplify notation, we write the off diagonal elements of the second order terms as $[\mathbf{D}_{pp}]_{12}=d_{pp}$, $[\mathbf{D}_{qp}]_{12}=d_{qp}$, and $[\mathbf{D}_{qq}]_{12}=d_{qq}$.
Some algebra shows that the matrices ${\mathbf Q}_q$ and ${\mathbf Q}_p$ can be parameterized
in the following manner:
\begin{equation}
{\mathbf Q}_q = 
\left(
\begin{array}{cc}
e^{i\phi}\sqrt{-d_{qq}^*(\alpha+d_{qp}^*)}   &  d_{qp}+\alpha^* \\
 -d_{qq}^*  &  - e^{-i\phi}\sqrt{-d_{qq}(\alpha^*+d_{qp})}
\end{array}
\right).
\end{equation}
\begin{equation}
{\mathbf Q}_p = 
\left(
\begin{array}{cc}
e^{i\phi}\sqrt{\alpha d_{pp}^*}   &  d_{pp} \\
 \alpha  &  - e^{-i\phi}\sqrt{\alpha^* d_{pp}}
\end{array}
\right),
\end{equation}
Here, $\phi$ is the phase of the coupling, $\tilde\eta = \vert \tilde\eta \vert e^{i\phi}$, and $\alpha$ is given by the solutions to the equation
\begin{equation}
\alpha^{*2}+d_{qp} \alpha^* +d_{pp} d_{qq} =0 .
\end{equation}
This transformation will make the off-diagonals in $\mathbf{D}'$ constants plus terms starting at $\mathcal{O}(\epsilon^3)$.  

In order to obtain the normal form of $\mathbf{D}$, we need to normalize the first order terms in the diagonals to have unit Poisson bracket.  We can interpret the new expressions at linear order
as new canonical variables $(q',p')$ if they are related to the old $(q,p)$ by a linear canonical
transformation:
\begin{equation}
\left(
\begin{array}{c}
q'  \\
p'     
\end{array}
\right)
=\frac{1}{\vert\mathcal{B}\vert^{1/2}}
\left(
\begin{array}{cc}
   1-2\Re (\alpha \tilde\eta^*) &  2\Re (d_1 \tilde\eta^*)   \\
  2\Re (d_3 \tilde\eta^*)  & 1+2 \Re (\alpha \tilde\eta^*)  
  \end{array}
\right)
\left(
\begin{array}{c}
q  \\
p     
\end{array}
\right) .
\end{equation}
Here the normalization is introduced to make $(q',p')$ a canonical pair.  It is given by the Poisson bracket of the transformed diagonal elements,
\begin{eqnarray}
\mathcal{B} &= \{D_{11}',D_{22}'\} = \left(1 - 4 (\Re (\alpha \tilde\eta^*))^2 - 4\Re (d_1 \tilde\eta^*)\Re (d_3 \tilde\eta^*) \right) \{q,p\} \\
&= \{q,p\} (1+\mathcal{O}(\epsilon^2)).
\end{eqnarray}

\section{Extension to Arbitrary Order\label{sec:extension}}

The calculation of \ref{sec:nf} can be extended to put the dispersion matrix into normal form order by order in powers of the phase space variables.  In this appendix, we outline the transformation, and show that there are enough free parameters in the near-identity transformations at each order to eliminate all non-constant terms in the off-diagonals elements of the dispersion matrix.
We still ignore Moyal corrections in this calculation.  Those will be discussed in the
next Appendix.

Start with the a dispersion matrix that is in normal form through order $N$.  This means that we can write the matrix as
\begin{eqnarray}\label{eq:expansion_of_D}
\mathbf{D}(q,p) = &\epsilon^0\mathbf{D}_{0}(q,p) + \epsilon^1\mathbf{D}_{1}(q,p) + \epsilon^2\mathbf{D}_{2}(q,p) + \epsilon^3\mathbf{D}_{3}(q,p) + \ldots \nonumber\\
& + \epsilon^N\mathbf{D}_{N}(q,p) + \epsilon^{N+1}\mathbf{D}_{N+1}(q,p) + \ldots,
\end{eqnarray}
where $\mathbf{D}_{0}$ is the constant coupling
\begin{equation}\label{eq:D0}
\mathbf{D}_0(q,p) = \left( \begin{array}{cc}
0& \tilde\eta \\
\tilde\eta^* & 0
\end{array}\right),
\end{equation}
and each of the matrices $\mathbf{D}_{j}(q,p)$, for $j\leq N$, are diagonal matrices with entries which are homogeneous polynomials of order $j$.  
At ${\mathcal O}(\epsilon^m)$ there are $m+1$ such matrices,
one for each monomial $q^{m'}p^{m-m'}$ for $m'=0,1,\ldots m$.

In particular, we can assume that the first order term has been transformed into the normal form
\begin{equation}\label{eq:D1}
\mathbf{D}_1(q,p) = \left( \begin{array}{cc}
-p& 0 \\
0 & q
\end{array}\right).
\end{equation}

We now want to apply a near-identity change of polarization basis which puts this matrix into normal form through order $N+1$.  Write the transformation matrix as
\begin{equation}\label{Q}
\mathbf{Q}(q,p) = \mathbf{1} + \epsilon^N\tilde{\mathbf{Q}}(q,p),
\end{equation}
where $\tilde{\mathbf{Q}}(q,p)$ is a matrix of homogeneous polynomials of order $N$.  
There are $N+1$ monomials of this order, hence there are $4(N+1)$ complex parameters
($8(N+1)$ real parameters) to work with.

The transformed dispersion matrix can then be written as
\begin{eqnarray}\label{Dprime}
\mathbf{D}' = \mathbf{D} + \epsilon^N\tilde{\mathbf{Q}}^\dagger\cdot \mathbf{D} + \epsilon^N\mathbf{D} \cdot\tilde{\mathbf{Q}} + \epsilon^{2N}\tilde{\mathbf{Q}}^\dagger\cdot \mathbf{D}\cdot \tilde{\mathbf{Q}}.
\end{eqnarray}
The last term in this expression starts at order $2N$.  Since we only need to consider terms of order $N$ and $N+1$, we can drop the last term except in the case $N=1$.  However, this is the case considered in \ref{sec:nf}, and so we do not need to consider it here.  We can now use Equation (\ref{eq:expansion_of_D}) to group the remaining terms in $\mathbf{D}'$ by their order;
\begin{eqnarray}
\mathbf{D}' &= \epsilon^0\mathbf{D}_{0} + \epsilon^1\mathbf{D}_{1} + \ldots + \epsilon^{N-1}\mathbf{D}_{N-1} 
& \mathcal{O}(<N) \\
&\quad + \epsilon^N(\mathbf{D}_{N} + \tilde{\mathbf{Q}}^\dagger\cdot \mathbf{D}_0 + \mathbf{D}_0 \cdot\tilde{\mathbf{Q}} )
& \mathcal{O}(N) \\
&\quad + \epsilon^{N+1}(\mathbf{D}_{N+1} + \tilde{\mathbf{Q}}^\dagger\cdot \mathbf{D}_1 + \mathbf{D}_1 \cdot\tilde{\mathbf{Q}} )
& \mathcal{O}(N+1) \\
&\quad +\ldots
& \mathcal{O}(>N+1) .
\end{eqnarray}
The order $N$ and $N+1$ terms are the ones which now need to put into normal form.  For this 1-dimensional problem, this means that the matrices
\begin{eqnarray}
	\tilde{\mathbf{Q}}^\dagger\cdot \mathbf{D}_0 + \mathbf{D}_0 \cdot \tilde{\mathbf{Q}}
\end{eqnarray}
and 
\begin{eqnarray}
	\mathbf{D}_{N+1} + \tilde{\mathbf{Q}}^\dagger\cdot \mathbf{D}_1 + \mathbf{D}_1 \cdot\tilde{\mathbf{Q}}
\end{eqnarray}
must be put into diagonal form.  Using Equations (\ref{eq:D0}) and (\ref{eq:D1}), we can simplify the equations for the elements of $\tilde{\mathbf{Q}}$.  The constraint on the order $N$ matrix gives 
\begin{eqnarray}\label{eq:order_N_eqn}
	[\tilde{\mathbf{Q}}]_{11} + [\tilde{\mathbf{Q}}^*]_{22} = 0.
\end{eqnarray}
These are $N+1$ complex ($2N+2$ real) conditions, one for each monomial of 
order $N$.  This leaves us with $(8N+8)-(2N+2)=6N+6$ free parameters.

The ${\mathcal O}(\epsilon^{N+1})$ normal form condition gives
\begin{eqnarray}\label{eq:order_N+1_eqn}
	p [\tilde{\mathbf{Q}}]_{12} - q [\tilde{\mathbf{Q}}^*]_{21} = [\mathbf{D}_{N+1}]_{12},
\end{eqnarray}
assuming that $\mathbf{D}_{N+1}$ is a hermitian matrix.  There are $N+2$ monomials of
order $N+1$.  Therefore, this set of constraints uses up only another $2N+2$ real parameters out of
our remaining allotment of $6N+6$. Therefore, generically, there is more than enough freedom to
carry out the normal form transformation, order by order.  We now consider how this picture
changes when we include Moyal corrections.

\section{Moyal Corrections for Phase Space Dependent Changes of Polarization\label{sec:moyal}}

In the previous section, a phase space dependent change of polarization is used to put the dispersion matrix into the form where its off diagonal elements are constants with a small perturbation that starts at order $\epsilon^3$.  This transformation is achieved through the conjugation
\begin{equation}
\mathbf{D}'(z) = \mathbf{Q}^\dagger(z) \cdot \mathbf{D}(z) \cdot \mathbf{Q}(z).
\end{equation}
However, since the matrices $\mathbf{D}$ and $\mathbf{Q}$ are actually matrix valued symbols of operators, we really need to use the Moyal star product to multiply the elements of the matrix.  The noncommutative star product is used in the symbol calculus so that the symbols (functions on phase space) maintain the commutation relations of the original operators.  So, we should actually use the expression
\begin{equation}
\mathbf{D}'(z) = \mathbf{Q}^\dagger(z) * \mathbf{D}(z) * \mathbf{Q}(z),
\end{equation}
where the matrices are multiplied in the usual way, but the elements of the matrices are multiplied using the star product.  In this section we will first consider how the Moyal corrections mix orders in $\epsilon$.  We will then use these results to sketch an argument that shows we can carry out the normal form transformation order by order, including Moyal terms.  All infinite series expressions should be interpreted formally.  We do not consider convergence, nor do we worry about the asymptotic character of these expressions.

The Moyal star product of two symbols $A(z)$ and $B(z)$ is often written as a formal power series (see \cite{Littlejohn} with $\hbar$ set to 1 since we are studying classical fields).  In general, if
the symbols are transcendental functions, this series will contain infinitely many terms:
\begin{eqnarray}\label{eq:moyal}
\fl A(z) \!* \!B(z) &= A(z) \exp\left( \frac{i}{2}  \frac{\overleftarrow \partial}{\partial z_\alpha} J_{\alpha\beta} \frac{\overrightarrow \partial}{\partial z_\beta} \right) B(z) \\
\fl & = A(z) \sum_{n=0}^\infty \frac{1}{n!}  \left( \frac{i}{2}  \frac{\overleftarrow \partial}{\partial z_\alpha} J_{\alpha\beta} \frac{\overrightarrow \partial}{\partial z_\beta} \right)^n  B(z) \\
\fl & = A(z) B(z) + \frac{i}{2} \{A,B\} -\frac{1}{8} \left( \partial^2_q A \, \partial^2_p B -2\, \partial_q\partial_p A \, \partial_q\partial_p B + \partial^2_p A \, \partial^2_q B \right)  + \ldots
\end{eqnarray}
\mycomment{%
\fl & = A(z) \cdot B(z) + \frac{i}{2} \{A,B\} -\frac{1}{8} \left( \{\partial_q A, \partial_p B \} -  \{\partial_p A, \partial_q B \}  \right)  + \ldots
\end{eqnarray}
}%
Here, $J_{\alpha \beta}$ is the symplectic matrix.  In the last line, the conjugate phase space coordinates $z=(q,p)$ are written explicitly to illustrate the nature of the terms in the previous sum.

We are assuming that the symbol of the dispersion matrix has a well defined Taylor's series in the mode conversion region, and that the lowest order terms in the series dominate.  This can be expressed mathematically by introducing the small parameter $\epsilon$.  Using the multi-index notation, we can expand the symbols in terms of all possible mononials:
\begin{equation}\label{eq:AB}
A(z) = \sum_{M=0}^\infty \epsilon^M \sum_{|m| = M} a_mz^m , \qquad
B(z) = \sum_{M=0}^\infty \epsilon^M \sum_{|m| = M} b_mz^m .
\end{equation}
The multi-index $m$ is a pair of integers $(m_1,m_2)$ with $|m|\equiv m_1 + m_2$, and $z^m$ defined as
\begin{eqnarray}
z^m = p^{m_1}q^{m_2}.
\end{eqnarray}
The terms $a_mz^m$ and $b_nz^m$ are homogenous polynomials of order $m$ in $z$. 
We will consider their dependence on $\epsilon$ momentarily.
The star product  of the series $A$ and $B$ is:
\begin{equation}\label{eq:Moyal-monomial}
A(z) * B(z) = \sum_{N,M=0}^\infty \epsilon^{M+N}  
    \sum_{|m| = M} \sum_{|n| = N} a_m b_n \, z^m * z^n .
\end{equation}
Now consider the star product of two generic monomials, $z^m * z^n$.  Because the star product can be written as a series in powers of derivatives, the star product $z^m * z^n$ contains powers of $z$ ranging
from $|l|_{max}=m+n$ to $|l|_{min}=|(m+n)-2min(m,n)|$.  All of the coefficients of this polynomial can 
be calculated from equation (\ref{eq:moyal}):
\begin{equation}
z^m * z^n = z^{m+n} + \sum_{|l|=||_{min}|}^{l_{max}-1} c_l(m,n) z^l .
\end{equation}
We will not need their explicit form, simply the fact that they are well defined functions of $(m,n)$ and
$l$.  It is also important to note that, because each term in the Moyal series~(\ref{eq:moyal}) involves 
derivatives acting both to the left and right, the series~(\ref{eq:Moyal-monomial}) descends in steps
of order $2$ in $z$ (e.g. $|m|,|m|-2,|m|-4,\ldots$).
This result is significant for us because it means that the star product only introduces monomials of degree less than the degree of the ordinary product.
We can use this result to write
\begin{eqnarray}
A(z) * B(z) &= A(z) B(z) + \sum_{N,M=0}^\infty \epsilon^{N+M}  
    \sum_{|m| = M} \sum_{|n| = N} a_m b_n  \sum_{|l|=0}^{L-1} c_l z^l     \\
&= A(z) B(z) + \sum_{L=0}^\infty \epsilon^{L}  
   \,  \sum_{|l|=0}^{L-1} c'_l z^l  \\
&= A(z) B(z) + \epsilon \left(\sum_{L=0}^\infty
 \sum_{|l|=0}^{L-1} \epsilon^{L-1-|l|}  \, c'_l \,  (\epsilon z)^l  \right)  .
\end{eqnarray}
This means that the original functional form we assumed, which ties the powers in $\epsilon$
directly to the order $z^{|m|}$ must be modified.  We replace it by the assumption that the
coefficient of $\epsilon^m$ has $z$-dependent coefficients that include no terms of higher order 
than $|m|$, but which can include all lower order terms in $z$. 
  It is clear that the Moyal corrections significantly complicate the algebra we have
to deal with.  In spite of this, it is still possible to show that we can arrive at well-posed formal 
iterations schemes for constructing these series.

As an example, consider the following problem: suppose we are given the symbol
$B(z)=1+\sum _{m=1}^\infty \epsilon^{|m|}b_m(z)$ where the $b_m(z)$ are independent of 
$\epsilon$ and include terms in $z$ of maximal order $|m|$. We now ask if we can find a symbol  
$A(z)$ of similar form such that $A*B=1$.
Write the formal series $A(z)$ as a Moyal product of terms:
\begin{equation}\label{eq:Aproduct}
A(z) = \ldots\left(1+\epsilon^3 a_3(z) \right)*\left(1+\epsilon^2 a_2(z) \right)*\left(1+\epsilon a_1(z) \right).
\end{equation}
Acting with $(1+\epsilon a_1(z))$ we have:
\begin{equation}
1=(1+\epsilon a_1(z))*\left(1 +\epsilon b_1(z) + \epsilon^2 b_2(z) + \epsilon ^3b_3(z)\ldots \right). 
\end{equation}
We collect terms in $\epsilon$ as usual, but now we must respect the Moyal product ordering:
\begin{equation}
1=1+\epsilon \left[a_1(z)+b_1(z)\right] + \epsilon^2 \left[b_2(z)+a_1(z)*b_1(z)\right] + \ldots . 
\end{equation}
This fixes $a_1(z)=-b_1(z)$ and determines all of the higher order terms, which now
become $b_n'(z)=b_n(z)-b_1(z)*b_{n-1}(z)$.  At next order, acting with $(1+\epsilon^2 a_2(z))$, we have:
\begin{equation}
1=(1+\epsilon^2 a_2(z))*\left(1  + \epsilon^2 b'_2(z) + \ldots \right). 
\end{equation}
This fixes $a_2(z)=-b'_2(z)$ and modifies all of the higher order even powers in $\epsilon$.
Clearly this procedure can be carried out to arbitrary order.  Using the insights gained from 
this simple example, we now return to the problem of casting $\mathbf{D}$ into normal form.

Once again, we assume that $\mathbf{D}(z)$ is in normal form up to ${\mathcal O}(\epsilon^{N})$.
To be precise: we assume that
\begin{equation}
\mathbf{D}(z) = \sum_{M=0}^\infty \sum_{|m|=M}\epsilon^m \mathbf{D}_m(z),
\end{equation}
where the terms up to $\epsilon^N$ are in normal form.
The $2\times 2$ hermitian matrices $\mathbf{D}_m(z)$ are now allowed to contain arbitrary terms in 
$z$ up to at most $\mathcal{O}(z^{|m|})$.  We use a congruence transformation 
$\mathbf{Q}$ of the same form as~(\ref{Q}), but we must now allow $\bar{\mathbf{Q}}$ to
be a general series in $z$ of terms up to order $|m|=N$. The calculation proceeds much as
before, but now the Moyal product is used instead of the ordinary product when two matrix
entries are multiplied. The terms up to order $N-1$ are unaffected.  At $\mathcal{O}(\epsilon^N)$
we have
\begin{equation}
\mathbf{D}'_N(z) = \mathbf{D}_N(z)
+\tilde{\mathbf{Q}}_N^*(z)\mathbf{D}_0+\mathbf{D}_0\tilde{\mathbf{Q}}_N(z).
\end{equation}
Because $\mathbf{D}_0$ has no $z$-dependence, this condition is unchanged from the 
case where Moyal corrections are ignored:
\begin{eqnarray}\label{eq:orderNeqn}
	[\tilde{\mathbf{Q}}_N]_{11}(z) + [\tilde{\mathbf{Q}}^*_N]_{22}(z) = 0.
\end{eqnarray}
At $\mathcal{O}(\epsilon^{N+1})$ we find now two conditions:
\begin{eqnarray}\label{eq:orderN1eqn1}
[\tilde{\mathbf{Q}}^*_N]_{21}*q-p*[\tilde{\mathbf{Q}}_N]_{12}   =- [\mathbf{D}_{N+1}]_{12},
\end{eqnarray}
and
\begin{eqnarray}\label{eq:orderN1eqn2}
q*[\tilde{\mathbf{Q}}_N]_{12} -[\tilde{\mathbf{Q}}^*_N]_{21}*p  =- [\mathbf{D}_{N+1}]_{12},
\end{eqnarray}
where we have used $[\mathbf{D}_{N+1}]_{12}=[\mathbf{D}_{N+1}]_{21}$ by hermiticity.
In addition, we require that the new diagonals at linear order in $q$ and $p$ are a conjugate
pair, which is one further condition.
At a fixed order N in $z$, there are $N+1$ monomials. Hence, because these expressions
include terms from $z^0$ up to order $N$, there are
\begin{equation}
\sum_{M=0}^N(M+1)=\frac{N^2+3N+3}{2},
\end{equation}
conditions that must be satisfied in each of these three equations~(\ref{eq:orderNeqn}), 
(\ref{eq:orderN1eqn1}) and~(\ref{eq:orderN1eqn2}).  This makes for $3(N^2+3N+3)$ real conditions
in total at order $N$.  But, there are $4(N^2+3N+3)$ parameters to work with (the four
entries of ${\bar Q}_N(z)$ each have $(N^2+3N+3)/2$ monomials with complex coefficients).  
Therefore, formally,
the normal form conditions can be satisfied order by order in $\epsilon$ and $z$ through a proper
choice of a $z$-dependent choice of polarization basis, including Moyal corrections.

\ack
The authors would like to thank A. N. Kaufman, for the inspiration and encouragement which stimulated this work, and for the insightful questions and comments during its revision.

This research was supported in part by an appointment to the U.S.~Department of Energy Fusion Energy Postdoctoral Research Program administered by the Oak Ridge Institute for Science and Education, the NSF-DOE Partnership in Basic Plasma Physics and the US DOE Office of Fusion Energy Sciences.

\section*{References}


\end{document}